\documentstyle[prb,eqsecnum,aps,twocolumn]{revtex}

\begin{document}
\twocolumn[{\hsize\textwidth\columnwidth\hsize\csname@twocolumnfalse%
\endcsname

\title{{\rm\small
 \rightline {UFR-HEP 00/21}
  \rightline {LMU-TPW 24/00}}
\vskip1truecm
 Fractional Quantum Hall Excitations as\\ RdTS Highest
Weight State Representations}
\author{I. Benkaddour $^{(1,2)}$, A. ELRhalami $^{(1)}$ and
E.H.Saidi$^{(1)}$}
\address{{ $^{(1)}$ High Energy Physics
Laboratory and UFR, Faculty of Sciences,\\Av Ibn Batota Po.Box
1014, Rabat, Morocco.\\
 $ ^{(2)}$ Sektion der Physik Theorie, Muenchen Universitat, Muenchen,\\Thereseinstr 37, D-80333, Muenchen,Deutschland.
} \\ \vspace{3mm} {\small e-mail:
ilham@theorie.physik.uni-muenchen.de\\arhalami@usa.net\\H-saidi@fsr.ac.ma}\\
\vspace{3mm}
\date{\today}} \maketitle

\begin{abstract}
Using the Chern-Simons effective model of fractional quantum Hall
(FQH) systems, we complete partial results obtained in the
literature on FQHE concerning topological orders of FQH states. We
show that there exists a class of effective FQH models having the
same filling fraction $\nu$, interchanged under $ Gl(n,Z)$
transformations and extends results on Haldane hierarchy. We also
show that Haldane states at any generic hierarhical level n may be
realised in terms of n Laughlin states composites and rederive
results for the n=2,3 levels respectively associated with $\nu=
{{2}\over{5}}$ and $\nu= {{3}\over{7}}$ filling fractions. We
study symmetries of the filling fractions series $\nu=
{{p_2}\over{p_1 p_2 -1}}$ and $\nu= {{p_1 p_2 -1}\over{p_1 p_2 p_3
-p_1 -p_2}}$, with $p_1$ odd and $p_2$ and $p_3$ even integers,
and show that, upon imposing the Gl(n,Z) invariance, we get
remarkable informations on their stability. Then, we reconsider
the Rausch de Traubenberg and Slupinsky (RdTS) algebra recently
obtained in [1,2] and analyse its limit on the boundary
$\partial({AdS_3})$ of the (1+2) dimensional manifold $AdS_3$. We
show that generally one may distinguish bulk highest weight states
(BHWS) living in $AdS_3$ and edge highest weight states
(EHWS)living on the border $\partial({AdS_3})$. We explore these
two kinds of RdTS representations carrying fractional values of
the spin and propose them as condidates to describe the FQH
states.
\end{abstract}
 } ]


\section{Introduction}
\qquad Recently a non trivial generalisation of the $(1+2)$
dimensional Poincar\'e algebra going beyond the standard
supersymmetric extension has been obtained in [1,2]. In addition
to the usual (1+2)d translation $P_{0,\pm}$ and rotation
$J_{0,\pm}$ symmetry generators, this extension refered herebelow
to as the Slupinski-Traubenberg algebra (RdTS algebra for short),
involves other kinds of conserved charges $Q_{s}$ and $\bar Q_{s}$
transforming as $so(1,2)$ Verma modules of spin $s=\pm {1\over k}
; {k\geq 2}$. RdTS fractional supersymmetry (fsusy) is a new
algebraic structure extending the standard structures in two
dimensions [3,4,5,6,7]. It involves a non abelian $SO(1,2)$
rotation goup, including the $SO(2)$ abelian subgroup appearing in
the usual constructions, and it is suspected to carry precious
informations on quantum physics in (1+2)dimensional systems with
boundaries; in particular in:\\ (${\bf a}$) classifying and
characterising the quasiparticles used in hierarchy building of
models of fractional quantum Hall effect (FQHE). \\ (${\bf b}$)
the study the propagation of strings on $AdS_3 \times N^{d-3}$;
the three dimensional anti de Sitter background times a compact
(d-3)-dimensinal manifold where d=26 for bosonic string and d=10
for type II superstrings.\\ Motivated by the following two:\\
(${\bf \alpha}$) similarities observed in [8] between RdTS highest
weight representations and quasiparticle states one encounters in
models of fractional quantum Hall liquids,\\ (${\bf \beta}$) exist
no consistent quantum field theoretical model yet of FQH liquids
leading naturally to FQH quasiparticle states; that is FQH
excitations carrying fractional values of the spin that are
generated by an (q-deformed) algebra of creation and annihilation
operators of certain quantum fields,\\ we want to explore in this
paper the issue of interpreting the FQH excitations as RdTS
highest weight states and the RdTS algebra as the algebra of
creation and annihilation operators of an hypothetic quantum field
model. In the present study we will also be interested in
analysing somes aspects on FQH hierarchies. In what follows we
propose to explicit a little bit our main motivations and the
purposes of our study :
\subsection{Main Motivations}
 In[8], we have studied aspects of the point (${\bf b}$) concerning the origin of RdTS symmetry and its link with space time boundary conformal invariance on $\partial{AdS_3}$. Here we want to analyse point (${\bf a}$) but also complete partial results obtained in the literature on FQHE regarding topological orders of FQH states. For these topological orders, one should note that they are generally classified by three rational numbers namely the filling fraction $\nu$, the electric charge $Q_e$ and the spin $s={\theta \over \pi}$ or equivalently by the hierarchical matrix $\bf{K}$, the charge vector $\bf{t}$ and the spin vector $\bf{S}$. As far as the filling fraction  $\nu$ is concerned, we want to show that there exists in fact a class of effective FQHE models having the same $\nu$ but not necessary the same kind of fundamental quasiparticles (elementary excitations). These models are interchanged by $ Gl(n,Z)$ transformations and turn out to englobe results on the leading orders of the Haldane hierarchy realised in terms of composites of Laughlin states. The $Gl(n,Z)$ symmetry appearing here is unusual, goes beyond the $Sl(n,Z)$ symmetry of the $\bf{Z^n}$  charge lattice and carries remarkable informations on the stability of the filling fraction states. \\
Before going into details, let us give other motivations supporting our interest in point (${\bf a}$). The basic idea is that in effective models of FQHE and RdTS algebra representations one has various quantities with striking analogies. In the Chern Simons (CS) effective model of FQHE, one encounters objects such as bulk states and edge states carrying fractional values of the spin as well as limits such the FQH droplet approximation allowing to make several estimations by using the conformal field theory (CFT) living on the border of the droplet. In the study of the RdTS algebra, one also has similar quantum states with the good features and, like for the droplet approximation of FQH liquids, one may here also make an analogous approximation by considering an appropriate (1+2)dimensional space time manifold $M$ with a boundary conformal invariance on $\partial M$. To have a more insight on this correspondence between FQH Chern-Simons effective field model and RdTS operators algebra, let us anticipate on some of our results by giving the following two:\\
($\bf i$) Standard representations of the RdTS algebra in (1+2) dimensions have quantum states carrying fractional values of the spin. However, for (1+2) dimensional space time manifolds  $M$ with a non zero boundary $\partial{M}$ and under a assumption to be specified later, the $ Q_{s}$ and $\bar Q_{s}$ charge operators may have different interpretations. They can be interpreted either as, $\bf {Q_{s}^+}$ and $ {\bf Q_{s}^-}$, generators of (1+2) dimensional quantum field states carrying fractional values of the spin or as $\bf {q_{s}^+}$ and $ {\bf q_{s}^-}$ and associated with highest weight states of a 2d boundary invariance. We shall refer to the first kind of states generated by $\bf {Q_{s}^+}$ and $ {\bf Q_{s}^-}$ as bulk highest weight states (BHWS for short) while the second ones generated by $\bf {q_{s}^+}$ and $ {\bf q_{s}^-}$ will be called edge highest weight states (EHWS). They live respectively in $M$ and $\partial M$.\\
($\bf ii$) In the CS effective model of FQHE, one also distinguish two kinds of states carrying fractional values of the spin: Bulk states and edge ones. The first ones are localised states; they are used in the (1+2)dimensional effective CS gauge theory in the study of hierarchies. The second states are extended states and are  described by a boundary conformal field theory in the droplet approximation. Edge states, which may also condensate, are responsible for the quantization of the Hall conductivity  $\sigma_{xy}$ as well as for the dynamics of the excitations on the boundary. \\

From this presentation, one sees that bulk (edge) excitations of the CS effective model of FQHE share several commun features with bulk (edge) representations of the RdTS algebra. They carry fractional spins and, up to an appropriate choice of the underlying geometry, are related to boundary conformal invariances living on the border of this geometry. One also learns that it may be possible to establish a correspondence between the CS effective field theory of FQH droplets and RdTS algebraic operators on $AdS_3$.
\subsection{Purpose and Presentation of the Work}
 The purpose of the present paper is to answer some of the questions rised above. It uses both quantum field theoretical methods and algebraic constructions and aims to reach two objectives:\\
 ($\alpha$) First we want to complete partial results obtained in the literature on FQHE concerning topological orders of FQH states. More precisely we want to study aspects of topological properties of a generic Haldane state[9] at hierarhical level n in terms of orders of composites of n Laughlin states. Relaxing the $SL(n,Z)$ symmetry of the $Z^n$ lattice of the charges of the CS $U(1)^n$ effective gauge theory of FQHE by allowing  $GL(n,Z)$ transformations, we show by explicit computations that states of filling fraction $\nu_H = K^{-1}_{1 1}$ in the level n of Haldane hierarchy are realised in terms of composites of n Laughlin states of filling fractions $\nu_j$; $j=1,\ldots,n$. We derive the general expression of the $\nu_j$ series and study their symmetries. We also give details for the two leading series $\nu= {{p_2}\over{p_1 p_2 -1}}$ and $\nu= {{p_1 p_2 -1}\over{p_1 p_2 p_3 -p_1 -p_2}}$, with $p_1$ odd and $p_2$ and $p_3$ even integers, and use the $Gl(n,Z)$ symmetries to derive informations on the stability of the filling fraction states.  \\
($\beta$) Second we examine the connection between RdTS representation states and those involved the CS effective theory of FQHE which we have described in point(ii) given subsection (1.1). Too particularly, we show that quasiparticles of fractional spins involved in the building of FQHE hierarchies could be interpreted as highest weight representation (HWR) states of the RdTS invariance. We also give arguments supporting a possible correpondence between FQH droplets and RdTS on $AdS_3$. \\
The presentation of this paper is as follows: In section 2, we review some general results of (1+2) dimensional effective CS model of the fractional quantum Hall liquids useful for our present study. We give the essential about FQHE features one needs  and recall the relevant properties of fractional quantum Hall states as well as topological orders and hierarchies. In subsection 2.2, we use the droplet approximation and we study the edge excitations of FQH droplets in terms of vertex operators living on the boundary of the droplet. In section 3, we reexamine the Haldane hierarchy of FQH droplets by taking into account the dynamics of the edge excitations as well as the interactions of branches. We first use appropriate $GL(n,Z)$ changes of CS gauge variables, going beyond the $SL(n,Z)$ symmety of the hypercubic $Z^n$ charge lattice, to reinterpret a generic level n Haldane state of filling fraction $\nu_H$ as a composite of n Laughlin states of filling fractions $\nu_i$; $i=1,\ldots,n$. We show amongst others that $ \nu_{H}={\sum_{i=0}^{n-1}} {1\over{m_{i}m_{i+1}}}$, where the $m_i$ integers are the solutions of the series $m_{i}=p_{i}m_{i-1}-m_{i-2}$; with $m_0=1$ and $m_1=p_1$ and have the property that the ${m_{i}m_{i+1}}$ product is usually an odd integer. We also discuss the effect of $Gl(n,Z)$ transformations on the Haldane series and the stability of the states. Then, we consider the branch interactions and show that the full hamiltonian is diagonalised by introducing effective velocities. In sections 4 and 5, we review the main lines of the construction of the RdTS algebra living in a (1+2) dimensional manifold with a boundary and show in section 6 why its HWR states may be viewed as condidates to describe bulk and edge excitations of the Chern Simons effective field theory of FQH liquids. In section 7, we make a discussion and give our conclusion.
\section{Generalities on FQH liquids.}
Roughly speaking, quantum Hall systems are defined as systems of electrons confined in a two dimensional layer embeded in a perpendicular external magnetic field B taken sufficiently strong so that electrons living on are totally polarized. The fractional Hall system should be incompressible [10], a feature which physically is interpreted as due to the existence of a positive energy gap for certain critical values of the filling fraction $\nu$. The latter is given by the number of electron $N_e$ divided by the number $N_{\phi_0}$ of unit of quantum flux $\phi_0$. For each critical value $\nu$, we have to distinguish localized and non localised states [13]. Localised states are bulk states described in the field approach by a (1+2)dimensional effective CS abelian gauge theory while non localised states are extended edge states described, in the droplet model, by a two dimensional boundary quantum field theory. Extended edge states are the carriers of the electric charge responsible of the non zero Hall conductivity  $\sigma_{xy}$ and its quantization [14]. The Hall current $I_h$ which is proportional to the filling fraction $\nu$ comes from the non localized states and is believed to arise from skipping orbits of cyclotronic electrons elastically scattered by the edge potential barrier [10]. Non localized states are generated by both bulk and edge elementary exitations carrying fractional spins  $s={\theta \over \pi}$ and fractional electric charges $Q$[11,15,16,17,18]. Non localized states condensate for some critical values of $\nu$ and it is admited that the Hall plateau of crtical value $\nu$ observed in experiments is indeed associated with the condensation of excitations. In FQH hierarchical models, we may have successive condensations of quasiparticles as in Haldane,Halperin, Jain and Zee hierarchies. We shall consider hereafter the Haldane hierarchy based on the Laughlin ground state of filling fraction $\nu={1\over m}$. For a review on the other hierarchies, see for instance[19,20,21,22 ].     \\
 At low energies, the effective field theory of the hierarchical FQH bulk states is described by an abelian CS gauge theory in $(1+2)$ dimensions. In this formulation, the fractional spin s appears as one of the parameters characterizing the topological orders in quantum Hall systems. The spin s, which arises generally from two sources, is given by the sum of two rational parameters ${\eta}$ and ${\theta}$ as shown herebelow:
 \begin{equation}
\pi s = {\theta}+{\eta}.
\end{equation}
 The first term, denoted by ${\theta}$ in eqs $(2.1)$, comes from the Bohm-Aharonov effect due to the presence of the magnetic field B while the second parameter ${\eta}$ is due to the Berry's phase induced by the curvature of the 2d space geometry of the system. Put differently, ${\eta}$ is related to the curvature R of the geometry of the space; it is non zero for the two sphere ${S^2}$ whereas it vanishes on the two plane $R^{2}$ and the two Torus $T^{2}$. For more details concerning the features of the ${\eta}$ term of eqs (2.1), see for example [11,23 ]. We shall focus our attention on just the first term ${\theta}$ by considering FQH models in $(1+2)$dimensions with a flat space geometry and study a new feature of those fractional quantum Hall excitations carrying fractional values of the spin and the charge.
\subsection{Chern-Simons model of FQH liquids.}
In the effective abelian CS gauge theory of the Haldane hierarchy of the FQH bulk states, the statistical angle ${\theta}$ is given by
\begin{equation}
\theta = {\ell}^{T}K^{-1}{\ell}=
{\sum}_{i}{\ell}^{i}(K^{-1})_{i}^{j}{\ell}_{j}
\end{equation}
In this eq, the $K_{ij}$ matrix has integer entries ( $K_{11}$ is odd integer while $K_{ii}$, $i\ge 2$, are even) and is intimately  related to the filling fraction $\nu$. The $\ell_i$ quantum numbers are topological charges of the elementary excitations of the FQH states. The latters are generally interpreted as quasiparticles or quasiholes, according to the sign of their electric charge $Q=-e\sum_i{K_{1i}l^i}$, and play an important role in the building of the Haldane hierarchy. To see the origin of the expression of the $\theta$ phase, we describe hereafter the main lines of the derivation of eq (2.1). To that purpose, consider a  polarized spin layer quantum Hall system in an external electromagnetic field B of potential $A_{\mu}{(x^{0},x^{1},x^{2})}$ with ${\mu}=0;1;2$. The interacting lagrangian, describing this strongly correlated electronic system in presence of B, is given by:
\begin{equation}
L= -e J_{\mu}A^{\mu}
\end{equation}
The effective theory of the Laughlin ground state at filling fraction $\nu ={1\over{m}}$ is described by an abelian U(1) CS gauge theory in (1+2) dimensions of gauge field $a_{\mu}$.
\begin{equation}
L= -[{m\over{2}}a_{\mu}\partial_{\nu}a_{\lambda}+eA_{\mu}\partial_{\nu}a_{\lambda}]{1\over{2\pi}}{\epsilon}^{\mu\nu\lambda}
\end{equation}
Eq $(2.4)$ may be obtained from eq (2.3) by using the hydrodynamic approach of the incompressible Hall liquid which takes the conserved electron current $J_\mu$ as:
\begin{equation}
{J_\mu}={1\over{2\pi}}\partial_{\nu}a_{\lambda}{\epsilon}^{\mu\nu\lambda}
\end{equation}
Note that the pure CS term in eq(2.4) endows each electron by m flux quanta as it may be seen from the eq of motion of the $a_0$ time component of the u(1) gauge field.
\begin{equation}
J^{\mu}=-{e \over {2\pi}}{1\over m}\epsilon^{\mu\nu\lambda}\partial_{\nu}A_\lambda
\end{equation}
 This eq describes the linear response of the ground state to the external magnetic field. Note also that up to boundary terms eq (2.4) is invariant under the U(1) gauge transformations,
\begin{equation}
a_{\mu}\rightarrow a'_{\mu}=a_{\mu}+{\partial}\lambda
\end{equation}
 A more complete effective description of the fractional quantum Hall systems is obtained by taking into account the elementary excitations effects. Introducing quasiparticle excitations of U(1) charge q in the effective theory (2.5) by inserting a source term type
\begin{equation}
qa_{\mu}J^{\mu}
\end{equation}
and evaluating the eq of motion of the time component of the gauge field $a_{\mu}$ by varying eqs (2.4), one discovers the two following: (i) the electric charge Q of the excitations is fractional as shown herebelow:
\begin{equation}
Q=-e{q\over m}
\end{equation}
For $q=1$ for instance, the fundamental quasiparticles have a fractional electric charge $-{e\over m}$, carries  $({1\over m})$ unit of the $a_{\mu}$-flux and an induced spin $s= {\theta \over \pi}$.  (ii) Considering two excitations carrying $q_{1}$ and $q_{2}$  charges moving one around the other, a Bohm-Aharonov phase $\phi$ = $2\pi {q_{1}q_{2}\over {m}}$ is induced. If moreover the two excitations are identical $q_{1}=q_{2}=q$, one gets then a statistical angle $\theta$ given by:
\begin{equation}
\theta=\pi{q^{2}\over m}
\end{equation}
 Note that $\theta\over\pi$is fractional and has to be distinguished from the intrinsic spin of the electrons. It is a topological parameter induced by the presence of the B magnetic field. Note also that the fundamental quasiparticles might be viewed too simply as elementary excitations of fermions. From this view point, a bound state of m elementary excitations of spin $ s={1\over m}$ and electric charge $Q=-e{1\over m}$ behave exactly like a fermion of electric charge -e. The above analysis extends straightforwardly for the description of the full Haldane hierarchy based on the Laughlin ground state. At the n-th level, $n>0$, the low energy effective theory of the Haldane hierarchy is described by an abelian $U^{n}(1)$ CS gauge theory in $(1+2)$dimension. The effective lagrangian of this model is given by.
\begin{equation}
L= -{1\over{4\pi}}{\epsilon}^{\mu\nu\lambda}[K_{ij}{a^{i}}_{\mu}\partial_{\nu}{a^{j}}_{\lambda}+2eA_{\mu}\partial_{\nu}(t_{i}{a^{i}}_{\lambda})]+{\ell}_{i}{a^{i}}_{\mu}J^\mu,
\end{equation}
where $K_{ij}$ are integers, $K_{11}$ odd and $K_{ii}$ even, characterizing the hierarchy, $t_{i}=\delta_{1i}$,is the so called vector charge and the $\ell_{j}$'s the number of quasiparticles of the $j^{th}$ hierarchical level. $K_{ij}$, $t_{i}$ and $l_{i}$ are quantum numbers describing three topological orders often represented by the filling fraction $\nu$ of the quantum hall liquid, the U(1) electric charge Q and the $\theta$ statistical angle of the quasiparticles. Following the analysis of [ 11,20,15] by taking the matrix K as,
\begin{equation}
K=\left(\matrix{
P_{1}&-1\cr
-1&P_{2}\cr
\\*&\\*&\ddots&-1\cr
\\*&\\*&-1&P_{n}
\cr}\right)
\end{equation}
where $p_{1}=m$ and $p_{j} = 2k_{j}$; j = 2,….,n, the quantities $\nu$, Q and $\theta$ are given by:
\begin{equation}
\begin{array}{lcr}
\nu ={\sum_{ij}}t^{i}{(K^{-1})_{i}}^{j}t_{j}\\
\\
Q=-e{\sum}_{ij}{t}^{i}(K^{-1})_{i}^{j}{\ell}_j\\
\\
\theta =\pi{\sum}_{ij}{\ell}^{i}(K^{-1})_{i}^{j}{\ell}_j
\end{array}
\end{equation}
Computing the inverse of eq(2.12) and setting $t_{i}=\delta_{1i}$, one gets for $\nu$ and Q.
\begin{equation}
\begin{array}{lcr}
\nu=K_{11}^{-1}={1\over{P_{1}-{1\over{P_{2}-{1\over{\ldots-{1\over P_{n}}}}}}}}\\
\\
Q=-e{\sum}_{j}(K^{-1})_{i}^{j}{\ell}_j
\end{array}
\end{equation}
At the second level of the hierarchy one obtains, by putting n=2 in the above eqs, the following:
\begin{equation}
\begin{array}{lcr}
{\nu}={P_{2}\over{P_{2}P_{1}-1}}\\
\\
Q=-e{{P_{2}\ell_{1}+\ell_{2}}\over{P_{2}P_{1}-1}}\\
\\
{\theta}={{\pi}\over{{P_{2}P_{1}-1}}}[P_{2}{\ell_{1}}^{2}+\ell_{1}{\ell_{2}}^{2}+2\ell_{1}\ell_{2}]
\end{array}
\end{equation}
In the end of this section, we would like note that in the above model, the effects of the boundary states are ignored. A complete study of the FQH liquids should however carry these effects as it is also required by experiments which reveal the existence of edge excitations with finite velocities. In the next section, we show how these effects may be incorporated. Later on we also examine their couplings.
\subsection{Edge excitations}
The edge excitations of the FQH liquids are conveniently described by a 2d boundary conformal field theory whose action may be obtained by evaluating eq(2.4) on the boundary of the system. The result one gets is:
\begin{equation}
S_{edge}={1\over 4\pi}{\int_{\partial M}}dtdx[K_{ij}\partial_{0}{\phi^{i}}\partial_{x}\phi^{j}-V_{ij}\partial_{x}\phi^{i}\partial_{x}\phi^{j}].
\end{equation}
To establish this relation, let us introduce some convention notations. (a) We denote by M the $(1+2)$ dimensional space-time, parameterized  by the local coordinates $(x^{0},x^{1},x^{2})$ in which the Chern-Simons gauge fields $a_{\mu}$ live and by $\partial M$ its boundary. (b) We write M as $M =R \times \Sigma$ where $\Sigma$ is the subset parameterized by the space variable $x^{1}$ and $x^{2}$, representing the surface where evolve the electrons of the FQH liquid. In this case, $\partial M$ is given by
\begin{equation}
\partial M=R \times \partial{\Sigma}
\end{equation}
where $\partial{\Sigma}$ stands for the one dimensional border of $\Sigma$. To get the boundary effective theory of the FQH states with finite velocities, one has to work a little bit hard. First one should note that the pure Chern-Simons gauge action,
\begin{equation}
S=-{1\over
4\pi}\int_{M}{a_{\mu}}^{i}[K_{ij}\partial_{\nu}{a_{\rho}}^{j}\epsilon^{\mu\nu\rho}]
\end{equation}
is, in general, not invariant under the gauge transformation of the CS gauge fields ${a_{\mu}}^{i}$:
\begin{equation}
{{a^{\prime}}_{\mu}}^{i}={a_{\mu}}^{i}+{\partial_{\mu}{\lambda}^{i}}  ;   i= 1,...,n,
\end{equation}
where $\lambda^{i} =\lambda^{i}(x^{0},x^{1},x^{2})$ are gauge parameters.  Putting the change (2.19) into eq (2.16), one obtains the following variation of the gauge action
\begin{equation}
\Delta S=-{1\over
2\pi}\int_{M}{\partial_{\mu}}[K_{ij}\lambda^{i}\partial_{\nu}{a_{\rho}}^{j}\epsilon^{\mu\nu\rho}]
\end{equation}
Integrating eq (2.20) by using Stokes theorem, we get by using eq (2.17)
\begin{equation}
\Delta S={1\over 2\pi}\int_{\partial
M}K_{ij}\lambda^{i}[\partial_{0}{a_{t}}^{j}-\partial_{t}{a_{0}}^{j}]
\end{equation}
where the index t refers to the tangent direction of $\partial \Sigma$.
In general eq (2.21) is non zero. To restore the gauge invariance of the action S eq(2.16), we require that gauge parameters $\lambda^{i}(x^{0},x^{1},x^{2})$ vanish on $\partial M= R \times \partial \Sigma$. In other words we demand the following:
\begin{equation}
\lambda (x^{0},\{x^{1},x^{2}\}\in \partial \Sigma)=0
\end{equation}
However there is a price one should pay for this choice since due to the restriction (2.22), some degrees of freedom of the ${a_\mu}^{i}$ gauge fields on the boundary become dynamical. These degrees of freedom give us the boundary effects we are looking for. To describe the dynamics of the boundary gauge degrees of freedom, we need to specify a gauge fixing condition. One way to do it is to take ${a_{0}}^{i}=0$ and interpret the bulk equations of the motion for ${a_{0}}^{i}$ as constraint eqs. Thus one remains with the eqs of motion  of ${a_{j}}^{i}$ which read as:
\begin{equation}
\epsilon^{0 \alpha\beta}[K_{ij}\partial_{\alpha}{a_\beta}^{j}]=0
\end{equation}
Eq (2.23) is solved by introducing a set of two dimensional $\{ \phi^i \}$ and taking the space components of the gauge fields${a_{\mu}}^{i}$ as:
\begin{equation}
{a_{\bf{j}}}^{i}=\partial_{\bf{j}}\phi^{i}
\end{equation}
Putting back this solution into eq (2.16) and again integrating by parts, one finds after using the choice ${a_{0}}^{i}=0$, the following:
\begin{equation}
S_{edge}={1\over 4\pi}\int_{\partial M}K_{ij}\partial_{0}\phi^{i}\partial_{t}\phi^{j}.
\end{equation}
This is not the full story since the action we have obtained has zero energy and so does not describe the boundary effects.
In fact, eq (2.25) recovers just a part of the result as it may be seen on eq(2.16). This ambiguity was expected because we knew already that edge excitations have finite velocities, a feature which has been ignored in the CS formulation of the FQH effective theory. To solve this problem, we have to find a way to insert the edge excitations velocities in the effective theory. A tricky way to do it is to put the velocities through the gauge fixing condition. For the simplicity of the presentation, we prefer to give hereafter a sketch of the method for the level one of the hierarchy where the $\bf{n \times n}$ matrix $K_{ij}$ reduces to the integer $P_{1}$ which we set m. This method is achieved in two steps: First make a special SO(1.2) Lorentz transformation on the $\{ x^{0}, x^{1}\}$ plane coordinates of the system $\{ x^{\mu}\}$
\begin{equation}
\begin{array}{ll}
y^{0}=x^{0}&;\partial{/}\partial y^{0}=\partial{/}\partial
x^{0}+v\partial{/}\partial x^{1}\\
y^{1}=x^{1}-vx^{0}&;\partial{/}\partial y^{1}=\partial{/}\partial
x^{1}\\ y^{2}=x^{2}&;\partial{/}\partial y^{2}=\partial{/}\partial
x^{2}
\end{array}
\end{equation}
where v is the velocity of the particle. Under this change, the gauge field $ a_{\mu}$ also transforms as
\begin{equation}
\begin{array}{lcr}
b^{0}=a^{0}+va^{1}\\ b^{1}=a^{1}\\ b^{2}=a^{2}
\end{array}
\end{equation}
while the pure U(1) CS action remains invariant as shown herebelow.
\begin{equation}
\begin{array}{lcr}
S&={m\over 4\pi}\int_{M}{d^{3}x}a_{\mu}\partial_{\nu}a_{\rho}\epsilon^{\mu\nu\rho}\\
\\
&={m\over
4\pi}\int_{M}{d^{3}y}b_{\mu}\partial_{\nu}b_{\rho}\epsilon^{\mu\nu\rho}
\end{array}
\end{equation}
The second step is to take $b_{0}=0$; that is $a_{0}= -v a_{1}$ instead of $a_{0}$=0 as done for deriving eq(2.26). Then integrating by part one finds, after renaming the tangent coordinate of $\partial\Sigma$ as x and its normal as y and using eqs(2.26), the following action
\begin{equation}
S={1\over 4\pi}\int_{\partial M}dtdx[m{{\partial\phi}\over\partial
t}{{\partial\phi}\over\partial x}+mv({{\partial\phi}\over\partial
x})^{2}]
\end{equation}
Eq (2.29) shows that the second term is just minus the hamiltonian of the edge excitation states and so the product mv should be negative definite. In other words consistency of the boundary effective theory requires $\it v$ and $\it m$ to have opposite signs. For a level $\bf n$ of the Haldane hierarchy, the effective edge theory is just an extension of the models (2.28) given by the action (2.16) where  $K_{ij}$ is a $\bf {n\times n}$ integer matrix defining the hierarchy and $V_{ij}$ is a positive definite matrix carrying the boundary effect of the FQH liquids. For a more rigorous derivation of the boundary effective action (2.16), see for instance [11,24].
\section{Fractional Quantum Hall droplet approach}
\hspace{1cm} So far we have learned that an abelian FQH system can
be described by the CS gauge theory eq(2.16) where the symmetric
matrix K defines topological order parameters characterizing the
internal structure of the system [25,26]. In the Haldane
construction on the Laughlin ground state, with filling fraction
$1\over m$, the electric charge $q_l$ of quasiparticles, the
statistical angle $\theta$ and the filling fraction of the
Hierarchical system are all of them expressed in terms of K, the
charge vector t and the $l_i$'s quantum numbers. Up on making an
assumption on the geometry of the FQH system approximating the
liquid to droplets, one can make several theoretical predictions
by using the power of the methods of CFT.\\ Here we want to use
the drolplet approach [27] of the FQHE in order to: (i) develop an
explicit derivation of a receipt giving a realisation of
hierarchical states at second level in terms of composites of two
Laughlin states. To our knowledge this special representation
which is often used in FQHE literature has never been prooved
rigorously. Here we not only answer this question but we give also
its extension to any level k of the Haldane hierarchy. More
precisely we show, by using an appropriate $GL(k,Z)$ change of CS
gauge variables, going beyond the $SL(k,Z)$ symmety preserving the
global property of the hypercubic $Z^k$ charge lattice, that we
can usually interpret a generic level k Haldane state of filling
fraction $\nu_H$ as a composite of k Laughlin states of filling
fractions $\nu_I$; $I=1,\ldots,k$. (ii) analyse the interactions
of the various droplet branch edge excitations by using the
special representation we refer to in the first point(i); see also
eq(5.6 ) and the subsequent one.  these couplings, which are
carried by a symmetric and positive matrix $V_{ij}$, may be
integrated out by making an appropriate choice of the CS gauge
field basis diagonalising the matrix $V_{ij}$. Like for known
results based on the representation(3.13 ), here also we show that
the full hamiltonian is diagonalised by introducing effective
velocities. (iii) build the wave functions of quasiparticles and
explore further the property of fractionality of their spins and
electric charges. These wave functions will be interpreted in
section 7 as HW states of RdTS representations.
\subsection{Haldane state as composites of Laughlin states}
To start, consider the effective bulk theory at a level k of
Haldane hierarchy described by CS gauge theory(2.11). Under a
specific linear combination of the CS gauge fields $a_\mu$ i.e,
\begin {equation}
\tilde{a}_{i \mu}= U_{ij}a_{j \mu},
\end {equation}
where $U_{ ij}$ is an invertible $k\times k$ matrix preserving the
charge quantization. $U_{ij}$ will be specified explicitly on the
levels $k=2$ and $k=3$ examples. Note that the $U_{ij}$ matrices
we will be considering here belong to $Gl(k,Z)$. They are not
symmetries of the CS gauge action (2.11) nor even of the $Z^k$
charge lattice in which the quantum topological vector t takes its
values. They are rather symmetries of the filling fraction $\nu_H$
eqs(2.13). Using the above transformation, one can diagonalise the
K matrix so that the effective lagrangian(2.11) is now replaced by
the new following one:
\begin{equation}
L=-{1\over
{4\pi}}\sum_{i}(\tilde{a}_{i\mu}D_{ii}\partial_{\nu}\tilde{a}_{i
\lambda}\epsilon^{\mu\nu\lambda} -{e\over
{2\pi}}T_{i}A_{\mu}\partial_{\nu}\tilde{a}_{i\lambda}\epsilon^{\mu\nu\lambda})
\end {equation}
where $D=U^{t}KU$. Eq(3.2) defines a system of k free abelian CS
gauge theories, where each one describes an effective model of a
Laughlin state of filling fraction $\nu_i =
{T_{i}}^{2}{(D^{-1})}_{ii}$. Integrating out the time component of
the gauge fileds $\tilde{a}_{i 0}$ by using equation of motion
${{\delta L}\over {\tilde{a}_{i 0}}}=0$, we find the following
filling fraction $\nu_{H}$ of the new system:
\begin{equation}
\nu_{H}=\sum_{i}{T_{i}}^{2}{(D^{-1})}_{ii}
\end{equation}
where the index H refers to Haldane. Now imposing invariance of
$\nu_H$ under the $Gl(k,Z)$ transformation,i.e;
\begin{equation}
\nu_{H}=\sum_{i}{T_{i}}^{2}{(D^{-1})}_{ii}={\sum_{ij}}t^{i}{(K^{-1})_{i}}^{j}t_{j}
\end{equation}
and choosing the T vector charge as $T=(1,\ldots,1)$, one can
determine explicitly the form of the U matrix of eq(3.1). Note in
passing that the above eq may be interpreted as describing the $
\nu_{H}$ state containing k components of incompressible fluids.
Each component $\nu_ i$,  $i =1,2,\ldots,k$, corresponds to a
Laughlin state with filling fraction $\nu_{j}={ 1\over{2m_{j}
+1}}$, $m_j$ integer, which in the hydrodynamical approach is
associated to a branch described by a c=1 free boundary CFT [11].
This interpretation is renforced by computing the total electric
charge and the spin of the full state. Mimicking the analysis of
section 2, we find after a straightforward calculations the two
following:
\begin {equation}
q_{\ell}= -eTD^{-1}\ell
\end{equation}
where $ \ell =(\ell_{1}...\ell_{k})$ is the toplogical charge of
the elementary excitations at a  $X_0$ position. Similarly we find
for the spin:
\begin {equation}
{\theta\over \pi}= \ell{D^{-1}}\ell
\end{equation}
To illustrate the idea let us give two examples; the first one
concerns a state at the second level of the Haldane hierarchy
(k=2) and the other one concerns a state at the third level of the
hierarchy(k=3). After these examples, we also give the general
result we have obtained.\\ {\bf(a)$\nu_{H}={2\over 5}$}\\
 Here we consider the FQH state with filling fraction $\nu_{H}={2\over 5}$; it has been studied from different views using too particularly Haldane, Halperin and Jain series [13,20]. In Haldane hierarchy, the $K_{2\over 5}$ matrix is:
 $$K_{2\over 5}=\left(\matrix{3&-1\cr -1 &2\cr}\right)$$
Next, choosing the U transformation matrix eq(5.1) as: $$U_{2\over
5}= \left(\matrix{1&1\cr 1&-2\cr}\right)$$ we get the following
diagonal matrix: $$D_{2\over 5}=\left(\matrix{3&0\cr
0&15\cr}\right)$$ So $\nu_{H}$ takes the remarkable decomposition,
$\nu_{H}= {1\over 3}+{1\over 15}$, which may viewed as a composite
configuration of two Laughlin fundamental states. This special
feature is renforced by the computation of the total electric
charge $q_{(\ell_1,\ell_2)}$ and the total statistics angle $
s_{(\ell_1,\ell_2)}$. For the electric charge we have the
following: $$q_{(\ell_1,\ell_2)}= -e({\ell_1\over3}+
{\ell_2\over15}).$$ Similarly  we have for the spin
$s_{(\ell_1,\ell_2)}={\theta\over \pi}$, $$s_{(\ell_1,\ell_2)} =
({{{\ell_1}^2\over3}}+ {{{\ell_2}^2\over15}}).$$
 Therefore the 2/5 FQH state may be interpreted as consisting of two free components of the incompressible fluid. The 2/5 FQH excitations consist then of two kinds of quasiparticles respectively given by the excitations of 1/3 and 1/15 Laughlin states. This analysis may be easily extended to the case of higher levels of the hierarchy as shown on the following example.\\
{\bf (b)$\nu_{H}={3\over7}$}\\ For the case of the third
hierarchical level at filling fraction $\nu_{H}={3\over7}$, where
the corresponding $K_{3\over7}$ matrix is:
  $$K_{3\over7}=\left(\matrix{3&-1&o\cr -1 &2&-1\cr0&-1&2}\right)$$
Moreover using the $U_{3\over7}$ matrix given by:
$$U_{3\over7}=\left(\matrix{1&1&1\cr 1 &1&-4\cr1&-2&-2}\right)$$
we get the following:
 $$D_{3\over7}=\left(\matrix{3&&\cr &15&\cr&&35}\right)$$
As discussed above for the level k=2 case, we have here also for
k=3; $$\nu_{H}={1\over 3}+{1\over 15}+{1\over 35},$$
$$q_{(\ell_1,\ell_2,\ell_3)}= -e({\ell_1\over3}+
{\ell_2\over15}+{\ell_3\over35}),$$ $$s_{(\ell_1,\ell_2,\ell_3)} =
({{{\ell_1}^2\over3}}+
{{{\ell_2}^2\over15}}+{{{\ell_3}^2\over35}}).$$ In summary the
$3\over7$ FQH state may be interpreted as a composite state
consisting of three branches with same propagating modes sign. We
shall return to study aspects of these propagating modes in a
moment; for the time being let us make two remarks. First note
that the $3\over7$ FQH state can be also viewed as a composite of
the $2\over5$ and the $1\over35$ FQH states in agreement with
associativity property of the tensor product of Hilbert space wave
functions. Second such analysis is a priori extendable to higher
orders of hierarchy. We have checked explicitly this feature for
the leading terms of the Haldane series and extended it to any
generic level k of the hierarchy using a special property of the
continuous fraction eq( 2.14). Indeed we have shown that (2.14)
can be usually put in the following remarkable form:
\begin{equation}
\nu_{H}={\sum_{j=0}^{k-1}} {1\over{m_{j}m_{j+1}}}
\end{equation}
where $m_{j}=p_{j}m_{j-1}-m_{j-2}$; with $m_0=1$ and $m_1=p_1$ and
where the $p_j$'s are as in eq (2.14). Note that the
${m_{j}m_{j+1}}$ product is usually a odd integer. Since $m_1$ is
odd it follows then all the $m_j$'s are odd and so the generic
terms $\nu_j= {1\over{m_{j}m_{j+1}}} $ may be interpreted as
filling fractions of laughlin states. From this eq, it is not
difficult to derive the $D^{k}_{\nu_H}$ which reads as:
\begin{equation}
D^{(k)}_{\nu_{H}}=\left(\matrix{m_{1} & & \cr & m_{1}m_{2} & \cr
& &\ddots & \cr  & & &m_{k-1}m_{k}\cr}\right)
\end{equation}
This eq establishes the extention to any value of k, the result we
have illustrated on the $\nu_H={2\over5}$ and $\nu_H={3\over7}$
examples. It shows clearly that a generic state at level k of the
Haldane hierachy may usually thought of as a composite of k
Laughlin states with different positive modes of branches. Note in
passing that our analysis is general as it also valid for branches
with opposite modes such as in the example $\nu_H={2\over3}$ state
belonging to the Haldane second level series by taking $p_2$ an
even negative integer.\\ Knowing $D_{\nu_{H}}$, one may also
derive the $ GL(k,Z)$ U-transformation by help of eq(2.12). Indeed
starting from the constraint eqs,
\begin{equation}
 (D_{\nu_{H}})_{ij}=U_{ni}(K_{\nu_{H}})_{nm}U_{mj}
\end{equation}
 where $ D_{ij}$ and $K_{nm}$ are respectively given by eqs (3.8 ) and (2.12 )  and where $U_{mj}$ are choosen as follows:
\begin{equation}
  U_{\nu_{H}}=\left(\matrix{1&1&\ldots&1\cr b_{21}&b_{22}&\ldots&b_{2k}\cr\vdots&\vdots&\ddots&\vdots\cr b_{k1}&b_{k2}&\ldots&b_{kk}\cr}\right)
\end{equation}
where $b_{ij}$ are integers to be determined. These eqs are non
linear numerical ones not easy to solve in general. In case of
Haldane hierarchy at the second level, eqs (3.9) has three
solutions given by:
\begin{equation}
\begin{array}{rcl}
 U_{(I)}&=&\left(\matrix{1&1\cr0&m_{1}\cr}\right);\\ \\
U_{(II)}&=&\left(\matrix{1&1\cr{2m_{1}\over
{m_{2}+1}}&m_{1}{({{1-m_{2}}\over{1+m_{2}}})}\cr}\right);\\ \\
U_{(III)}&=&\left(\matrix{1&1\cr{2m_{1}\over
{m_{2}+1}}&m_{1}\cr}\right)
\end {array}
\end{equation}
These solutions lead to the same $D^{(2)}_{\nu_{H}}$ matrix
\begin{equation}
D^{(2)}_{\nu_{H}}=\left(\matrix{m_{1}&0\cr 0&m_{1}m_{2}\cr}\right)
\end{equation}
but not all of them belong to $ GL(2,Z)$ . If, in addition to
$U_I$, we require that the two other transformations also belong
to $Gl(2,Z)$; that is $l(m_2 +1)= m_1$ for integer l's, we get
constraints on the FQH filling fractions. In this case the three
$U_{I}, U_{II} and U_{III}$ transformations are all of them in
$Gl(2,Z)$ and have a remarkable interpretations. They lead to
stable states associated with the observable filling fractions as
$v= 2/3, 2/5, 2/9,...$. For $l(m_2 +1)\neq m_1$, $U_{II} and
U_{III}$ are no loger acceptable since only $U_I$ is in $Gl(2,Z)$
. At first sight, the $\nu_H$ solutions allowed by $Gl(2,Z)$ turn
out to be associated with unstable states. We still do not
understand this property.\\ For level three of the hierarchy, eqs
(3.9) read as:
\begin{equation}
\begin{array}{lcr}
m_{1}-2a + a^{2}{{m_{2}+1}\over {m_{1}}} -2ad +
d^{2}{{m_{3}+m_{1}}\over{ m_{2}}}&=& m_{1}\\ m_{1}-2b +
b^{2}{{m_{2}+1}\over {m_{1}}} -2be + e^{2}{{m_{3}+m_{1}}\over{
m_{2}}}&=& m_{1}m_{2}\\ m_{1}-2c + c^{2}{{m_{2}+1}\over {m_{1}}}
-2cf + f^{2}{{m_{3}+m_{1}}\over{ m_{2}}}&=& m_{2}m_{3}\\
m_{1}-(a+b) +a b{{m_{2}+1}\over {m_{1}}} -ea-bd +
ed{{m_{3}+m_{1}}\over{ m_{2}}}&=& 0\\ m_{1}-(a+c) +a
c{{m_{2}+1}\over {m_{1}}} -fa-cd + fd{{m_{3}+m_{1}}\over{
m_{2}}}&=&0\\ m_{1}-(b+c) +b c{{m_{2}+1}\over {m_{1}}} -ce-fb +
ef{{m_{3}+m_{1}}\over{ m_{2}}}&=&0
\end{array}
\end{equation}
where we have taken $U_{ij}$ as,
\begin{equation}
U_{\nu_{H}}=\left(\matrix{1&1&1\cr a&b&c\cr d&e&f\cr}\right)
\end{equation}
Here also we have several solutions, an acceptable one of them was
already encountered previously eq (3.13 ); an other one is given
by the following triangular matrix.
\begin{equation}
U_{\nu_{H}}=\left(\matrix{1&1&1\cr 0&m_1&m_1\cr
0&0&m_2\cr}\right)\nonumber
\end{equation}
The above calculations may be extended for higher hierarchical
levels k ; the number of solutions rise assymptoticaly with the
level value of k. A particular solution for the generic case was
already identified, but still needs more analysis. Details on this
issue as well as branches with opposite modes and more
informations on the $Gl(k,Z)$ transformations will be reported
elsewhere [27 ].
\subsection{Dynamics of droplets edge excitations }
To start recall that in the droplet approximation, droplet waves
are identified as edge excitations of FQH fluid. These waves move
along the edge of the sample, considered here as a finite FQH
liquid with filling fraction $\nu= {1\over m}$ confined by a
smooth potential well, in which the electrons propagate with the
velocity $v={E\over B}c$ [28].\\ In the language of 2d CFT, edge
excitations of FQH liquid with k branches are described a c=k CFT
theory with a $U(1)^k$ Kac-Moody symmetry. For the leading case,
the corresponding c=1 CFT describes a Laughlin state having one
branch edge excitations and a disk like geometry. The particle
density $\rho$ is related to the bosonic field $\phi$ as  $\rho
(x) ={1\over 2\pi}\partial_{x}\phi$; this is just the electron
density operator on the edge which may be thought of as the U(1)
Kac-Moody current. Expanding this current in Laurent series in
terms of the Laurrent modes $\rho_{k}$ and using the U(1)
Kac-Moody algebra,
\begin{equation}
[\rho_{k},\rho_{k\prime}]={\nu\over{2\pi}}k\delta_{k+k\prime}
\end{equation}
one can show that the hamiltonian H giving the quantum energy of
the edge excitations is proportional to the velocity v and reads
in terms of the $\rho_{k}$ modes as:
\begin {equation}
\begin{array}{rcl}
H&=&-2\pi\sum_{k > 0}v \rho_{k}\rho_{-k},\\
\left[H,\rho_{k}\right]&=&-vk\rho_{k}
\end{array}
\end{equation}
Charged excitations are created by the following vertex field
operator $\psi$;
\begin {equation}
\psi \propto e^{-{1\over\sqrt{ \nu}}\phi}
\end{equation}
Moreover using the fact that in this CFT, $\phi$ is just a free
phonon field having a two point function $<\phi (z_1)\phi(z_2)>
\sim  lnz_{12}$, one can easily establish that the $\psi$ field
operator carries a unit electric charge and a conformal spin $s=
{1\over{2\nu}}$ as shown on the following commutation relations.
\begin{equation}
\begin {array}{rcl}
\left[\rho(x), \psi (x\prime)\right]&=& \delta (x-{x\prime})\psi
({x\prime})\\ \left[L_0, \psi (x)\right]&=& {1\over{2\nu}}\psi
({x}).
\end{array}
\end{equation}
 Similar calculations shows moreover that the propagator of the $\psi$ field operator is that of a chiral Luttinger liquid [29]. \\
$$ G(x,t) \propto {1\over ({x-vt})^{1\over \nu}}.$$ Hierarchical
states of level k  contain several component of the incompressible
fluid, each component give rise to a branch of the edge
excitations and so are described by a c=k CFT with $U^{k}(1)$ Kac
Moody symmetry.\\ Extending the previous analysis to generic k
values, we get the following relations regarding the $\rho_{i,k}$
field operators.
\begin{equation}
\begin{array}{rcl}
[\rho_{i,k}, \rho_{i,k\prime}]&=& T_{i}{(D^{-1})}_{i,j}{1\over
2\pi}k\delta_{k+k\prime},\\
 \rho_{e}&=&-e\sum_{i}\rho_{i},\\
\left[\rho(x), \psi (x\prime)\right]&=&\ell_{i}(D^{-1})_{ij}
\delta (x-{x\prime})\psi_{\ell} ({x\prime}),\\
\psi_{\ell}&\propto& e^{i{\ell_{i}\over T_{i}}\phi_{i}}
\end{array}
\end{equation}
In these eqs, the  $\phi_{i}$'s are the bosonic scalars associated
with each branch of the fluid while $\rho_{e}$ is the total
electron density. An electron excitation appears whenever $
\sum_{i}{(D^{-1})}_{ii}\ell_{i}=1$ is fulfilled; this implies in
turn that  $\ell_{i}=\sum_{i}D_{ij}L_j$ with $\sum_{i} L_{i}=1$.
The electron operator $ \psi_{e L}$ takes then the form:
\begin {equation}
\psi_{e,L}\propto e^{i\sum_{i}{\ell_{i}\over T_{i}}\phi_{i}}
\end{equation}
and obeys the following anticommutation relation:
\begin{equation}
\psi_{e L}(x)\psi_{e L}({x\prime})={(-1)}^{\lambda}\psi_{e
L}({x\prime})\psi_{e L}(x)
\end{equation}
where $\lambda =\sum_{i}L_{i}\nu^{-1}_{i}L_i$; i.e an odd integer.
Before going ahead note that each factor $ e^{i{\ell_{i}\over
T_{i}}\phi_{i}}$ in eq(3.20) defines the wave function of a
quasiparticle of fractional spin.
If one forgets about FQHE for a wile and just retains that on
$\partial(M)$ lives a conformal structure, one may consider its
highest weight representations which read in general as:
\begin{equation}
\begin{array}{lcr}
{L _{0}}\vert {h,{\bar{h}}}\rangle=h {\vert
{h,{\bar{h}}}\rangle},\\ {L_n}{\vert {h,{\bar{h}}}\rangle}=0;\quad
n\geq 1\\ {\bar L_{0}}{\vert {h,{\bar{h}}}\rangle}=\bar{h}{\vert
{h,{\bar{h}}}\rangle},\\ {{\bar{L}}_{n}}{\vert
{h,{\bar{h}}}\rangle}=0;\quad n{\geq}1,\\ cI\vert
{h,{\bar{h}}}\rangle=c\vert {h,{\bar{h}}}\rangle
\end{array}
\end{equation}
where $\vert {h,{\bar{h}}}\rangle$ are Virasoro primary states. To
make contact with the FQH quantities we have been considering, we
give the following correspondence. (i) the vertex operator
$e^{i{\ell_{i}\over T_{i}}\phi_{i}}$ may be represented generally
by a 2d field operator $\phi_{h,\bar{h}}(z,\bar z)$. This is a
primary conformal field representation of conformal scale
$\delta$= $h$+ $\bar{h}$ and conformal spin $s$= $h$- $\bar{h}$.
In the present case the primary fields are chiral and so
$\bar{h}=0$. (ii) The highest weight states $\vert{s}\rangle$
associated with the chiral fields $\phi_{s}(z)$ are related as
$\Phi _{s}(0){\vert{0}\rangle} =\vert{s}\rangle$ where now the
spin s is given by eqs(3.9) and (3.12).
\subsection{ Branch Interactions}
Recall that in absence of interactions and ignoring the edge
excitations velocities $v_{I}$, the action of the system reads as,
\begin{equation}
S=-{1\over 4\pi}\int{d^{3}x
D_{ii}{\tilde{a}}_{i\mu}\partial_{\nu}{\tilde{a}}_{i\lambda}\epsilon^{\mu\nu\lambda}}
\end{equation}
On the n branches of the droplet, the dynamics of the excitations
are conveniently described by a boundary c=n CFT. In what follows
we use the representation (3.19) in order to study the branch
interactions in presence of velocities $v_{i}$. We will show that,
up on redefining the Kac-Moody currents $\rho_{i}$, the FQH
droplet is still described by a c=n CFT provided replacing the
$v_{i}$'s by new velocities $\tilde{v_{i}}$. To that purpose we
will start first by introducing the $v_{i}$ velocities in the
system. Using the change(3.1-9) and following the same calculus we
have developed in section 2 by taking
$\tilde{a}^{0}_{i}=-v_{i}\tilde{a}_{i}^1$, we find, after
performing an integration by part, the following edge action:
\begin{equation}
S_{edge}={1\over
4\pi}\int{dxdt[D_{ii}\partial_{t}\phi^{i}\partial_{x}\phi^{i}+
v_{i}D_{ii}\partial_{x}\phi^{i}\partial_{x}\phi^{i}]}
\end{equation}
The edge dynamics is now described by the following hamiltonian
generalising eq(3.17);
\begin{equation}
H=-{1\over
2\pi}\sum_{i}v_{i}D_{ii}\partial_{x}\phi^{i}\partial_{x}\phi^{i}
\end{equation}
In this formalism we have here also assumed a large gap between
different edges; i.e  the radii $ r_i$ and $r_j$ of two droplets
are such that: $r_{i}-r_{j}\gg \ell_{o}^{2}$ where $ \ell_{o}$ is
the magnetic lenght. If now we take edges interactions into
account, the above Hamiltonian becomes:
\begin{equation}
H= {-1\over
2\pi}\sum_{i}V_{ij}\partial_{x}\phi^{i}\partial_{x}\phi^{j}
\end{equation}
where the symmetric $k\times k$ matrix $V_{ij}$ contains non
diagonal terms in addition to the diagonal ones eq(3.26 ). A way
to deal with this hamiltonian is to diagonalise $V_{ij}$ by
performing a transformation on the Kac-Moody currents
$\partial_{x}\phi^{i}$ and keeping $D_{\nu}$ diagonal. This gives
the new velocities $\tilde{v_{i}}$. Thus, under a linear
trasformation $W$, which act on $\rho_{i}$ as,
\begin {equation}
\rho_{i}=W_{ij}\tilde{\rho}_j
\end {equation}
we should have:
 $$W^{t}DW=D^{\prime}$$
 $$W^{t}VW=\lambda_{i}\delta_{ij}$$
 where $\lambda_{i}$ are real parameters defining the new velocities $\tilde{v_i}$. To illustrate the method, let us consider the case of FQH system with two branches and an interacting matrix.
$$ V=\left(\matrix{V_{11}&V_{12}\cr V_{12}&V_{22}\cr}\right).$$
 Under the particular transformation W,
\begin{equation}
\begin {array}{rcl}
\rho_{1}&=&{1\over
\sqrt{B_{1}}}cos{\alpha}\tilde{\rho}_{1}+{1\over
\sqrt{B_{1}}}sin{\alpha}\tilde{\rho}_{2}\\ \rho_{2}&=&-{1\over
\sqrt{B_{2}}}sin{\alpha}\tilde{\rho}_{1}+{1\over
\sqrt{B_{2}}}cos{\alpha}\tilde{\rho}_{2}
\end {array}
\end{equation}
where $B_{I}=T_{I}{v_{I}\over \nu_{I}}$, the V potential matrix is
diagonalized for $\alpha$ angles constrained as:
\begin{equation}
tg{2\alpha}= -2 {V_{12}\sqrt{B_{1}B_{2}}\over
{B_{2}V_{11}-B_{1}V_{22}}}
\end{equation}
 The new velocities are then,
\begin{equation}
\begin{array}{rcl}
\tilde{v}_{1}&=&{V_{11}\over B_{1}}{cos^{4}\alpha \over cos
2\alpha}-{V_{22}\over B_{2}}{sin^{4}\alpha \over cos 2\alpha}\\
\\
\tilde{v}_{2}&=&-{V_{11}\over B_{1}}{sin^{4}\alpha \over cos
2\alpha}+{V_{22}\over B_{2}}{cos^{4}\alpha \over cos 2\alpha}
\end{array}
\end{equation}
Finally, the diagonalised hamiltonian in terms of the new
variables as,
\begin{equation}
H=-{1\over
4\pi}\sum_{i}{\tilde{v_{i}}\partial_{x}{\tilde{\phi^{i}}}\partial_{x}{\tilde{\phi^{i}}}}
\end{equation}
and describes indeed a c=n boundary CFT. Note in the end of this
paragraph that W taken here is a $ O(k,R)$ transformation and so
affects the level of the $U(1)^k$ Kac Moody algebra eq(3.19).
Under the linear transformation W choosen as in eqs(3.28-29), the
new current operators $\tilde{\rho_i}$ generate a $U(1)^k$ Kac
Moody algebra of level one rather than a level $\nu$ as in
eq(3.19). Setting $B_{1}=1$ and $B_{2}=1$, the transfomation W
becomes orthogonal and so ${\rho_i}$ and $\tilde{\rho_i}$ obey the
same algebra.
\section{RdTS supersymmetry}
RdTS supersymmetry is a special generalisation of fractional
supersymmetry in two dimensions which was considered in many
occasions in the past in connection with integrable deformations
of conformal invariance and representations of the universal
envelopping $U_q{sl(2)}$ quantum ordinary and affine
symmetries[30,3,4]. Like for fsusy, highest weight representations
of RdTS algebra carry fractional values of the spin and obey more
a less quite similar eqs. Both fsusy and RdTS invariances describe
residual symmetries which are left after integrable deformations
of infinite dimensional invariances by relevant operators. We
propose to describe hereafter the main lines of RdTS invariance by
considering a (1+2) dimensional space time $M$ with a boundary
$\partial M $.
\subsection{Extension of the $\bf P_{(1,2)}$ Poincar\'e algebra}
To start consider the Poincar\'{e} symmetry of the $R^{1,2}$ space
generated by the space time translations $P_{\mu}$ and the Lorentz
rotations $J_{\alpha}$ satisfying altogether the following closed
commutation relations:
\begin {eqnarray}
\lbrack {J_{\alpha},J_{\beta}} \rbrack &=&i{\epsilon_{\alpha \beta
}\eta^{\gamma \delta} J_ {\delta}}\nonumber\\ \lbrack
{J_{\alpha},P_{\beta}}\rbrack &=&i{\epsilon_{\alpha \beta
\gamma}\eta^{\gamma \delta}P_ {\delta}}\\ \lbrack {P_{\mu},
P_{\nu}}\rbrack&=&0\nonumber
\end {eqnarray}
In these eqs, $\eta _{\alpha \beta}=diag (1, -1, -1)$ is the
$R^{1,2}$ Minkowski metric and $\epsilon_{\alpha \beta \gamma}$ is
the completly antisymmetric Levi-Civita tensor such that $\epsilon
_{012}=1$. A convenient way to handle eqs(4.1) is to work with an
equivalent formulation using the following Cartan basis of
generators $P_{\mp}=P_1{\pm}i P_2$ and $J_{\mp}=J_1{\pm}i J_2$. In
this basis eqs(4.1) read as:
\begin{eqnarray}
\lbrack {J_+, J_-}\rbrack &=& -2J_0 \nonumber\\ \lbrack {J_0,
J_{\pm}}\rbrack &=&  \pm {J_{\pm}}\nonumber\\ \lbrack {J_{\pm},
P_{\mp}}\rbrack &=& \pm{P_0}\\ \lbrack {J_+, P_+}\rbrack &=&
\lbrack {J_-, P_-}\rbrack = 0 \nonumber\\ \lbrack {J_0,
P_0}\rbrack &=& \lbrack {P_{\pm}, P_{\mp}}\rbrack = 0\nonumber
\end{eqnarray}
The algebra (4.1-2) has two Casimir operators $P^{2}= {P_0}^2 -
{1\over 2}(P_{+}P_{-} +P_{-}P_{+})$  and $P.J= P_{0}J_{0}-
{1\over2}(P_{+}J_{-} +P_{-}J_{+})$. When acting on highest weight
states of mass m and spin s, the eigenvalues of these operators
are $m^2$ and $ms$ respectively. For a given s, one distinguishes
two classes of irreducible representations: massive and massless
representations. To build the $so(1,2)$ massive representations,
it is convenient to go to the rest frame where the momentum vector
$P_{\mu}$ is $(m,0,0)$ and the $SO(1,2)$ group reduces to its
abelian $SO(2)$ little subgroup generated by $J_0$; $(J_{\pm}=0)$.
In this case, massive irreducible representations are one
dimensional and are parametrized by a real parameter. For the full
$SO(1,2)$ group however, the representations are either finite
dimensional for ${\vert {s} \vert}\in {\bf {Z}^{+}}/2$ or infinite
dimensional for the remaning values of $s$.\\
 Given a primary state $\vert {s} \rangle $ of spin $s$, and using the abovementioned $SO(1,2)$ group theoretical properties, one may construct in general two representations HWR(I) and HWR(II) out of this state $\vert {s} \rangle$. The first representation HWR(I) is given by:
\begin{eqnarray}
J^0{\vert {s}\rangle}&=& s{\vert {s}\rangle}\nonumber\\
J_{-}{\vert {s}\rangle}&=& 0\nonumber\\ {\vert {s,n} \rangle}&=&
\sqrt{{\Gamma (2s)\over{\Gamma(2s+n)\Gamma(n+1)}}}(J_{+})^n{\vert
{s}\rangle}, n{\geq}1\\
J_{0}{\vert{s,n}\rangle}&=&(s+n){\vert{s,n}\rangle}\nonumber\\
J_{+}{\vert{s,n}\rangle}&=&\sqrt
{(2s+n)(n+1)}{\vert{s,n+1}\rangle}\nonumber\\
J_{-}{\vert{s,n}\rangle}&=&\sqrt
{(2s+n-1)n}{\vert{s,n-1}\rangle}\nonumber
\end{eqnarray}
The second representation HWR(II) is defined as:
\begin{eqnarray}
\bar{J _0}{\vert {\bar{s}} \rangle} &=& -s {\vert
{\bar{s}}\rangle}\nonumber\\
 \bar{J_{+}}  {\vert {\bar{s}}\rangle}&=& 0\nonumber\\
{\vert{\bar{s},n}\rangle}&=&{(-)^n}\sqrt{\Gamma(2s)\over{\Gamma(2s+n)\Gamma(n+1)}}
(\bar{J_{-}})^{n}{\vert{\bar{s}}\rangle}\\ \bar{J _0}{\vert
{\bar{s},n} \rangle} &=& -(s+n) {\vert
{\bar{s},n}\rangle}\nonumber\\
\bar{J_+}{\vert{\bar{s},n}\rangle}&=&{-\sqrt{(2s+n-1)(n)}} \vert
\bar{s},n+1\rangle\nonumber
\end{eqnarray}
Note in passing that in the second module we have supplemented the
generators and the representations states with a bar index. Note
moreover that both HWR(I) and HWR(II) representations have the
same $so(1,2)$ Casimir $C_s$= $s(s-1), s<0$. For $s\in \bf Z^-/2$,
these representations are finite dimensional and their dimension
is $(2\vert {s} \vert+1)$. For generic real values of $s$, the
dimension of the representations is however infinite. If one
chooses a fractional value of $s$ say $s=-{{1}\over k}$; each of
the two representations (4.3-4) splits a priori into two
isomorphic representations respectively denoted as $D_{\pm 1/k}^+
and D_{\pm 1/k}^-$. This degeneracy is due to the redundancy in
choosing the spin structure of $\sqrt {-{2/k}}$ which can be taken
either as $+i\sqrt {{2/k}}$  or $-i\sqrt {{2/k}}$. These
representations are not independent since they are related by
conjugations; this why we shall use hereafter the choice of [1] by
considering only
 $D_{-{1/k}}^{+}$ and $D_{-{1/k}}^-$. In this case the two representation generators $J_{0,\pm}$ and $\bar{J}_{0,\pm}$ are related as:
\begin{equation}
\bar{J}_{0,\mp}=(J_{0,\pm})^*
\end{equation}
Furthermore taking the tensor product of the primary states
${\vert {s} \rangle}$ and ${\vert {\bar{s}} \rangle}$ of the two
$so(1,2)$ modules HWR(I) and HWR(II) and using eqs(4.3-4), it is
straightforward to check that it behaves like a scalar under the
full charge operator $J_0+{\bar{J}} _0$ :
\begin{equation}
(J_0+\bar{J}_0){\vert{s}\rangle}\otimes{\vert{\bar{s}}\rangle}= 0
\end{equation}
Eq(4.6) is a familiar relation in the study of primary states of
Virasoro algebra. This equation together with the mode operators
$J_{-}^n $ and  $\bar{J}_{+}^{m}$ which act on
${\vert{s}\rangle}\otimes{\vert{\bar{s}}\rangle}$ as:
\begin{eqnarray}
(J_{-})^n{\vert{s}\rangle}\otimes{\vert{\bar{s}}\rangle}&=&
0,\quad n{\geq}1\nonumber\\ (\bar{J}_{+})^m
{\vert{s}\rangle}\otimes{\vert{\bar{s}}\rangle}&=& 0,\quad
m{\geq}1
\end{eqnarray}
define a highest weight state which looks like a Virasoro primary
state of spin 2s and scale dimension $\delta= 0 $. In [8], we have
shown that eqs(4.6-7) are indeed related to HWR of conformal
invariance which is written as:
\begin{eqnarray}
({L _0}-{\bar{L} _0}){\Phi _{h,\bar{h}}}(0,0) {\vert{0}\rangle}&=&
(h-{\bar{h}}){\Phi _{h,\bar{h}}}(0,0) {\vert{0}\rangle}\nonumber\\
({L _0}+{\bar{L} _0}){\Phi _{h,\bar{h}}}(0,0) {\vert{0}\rangle}&=&
(h+{\bar{h}}){\Phi _{h,\bar{h}}}(0,0) {\vert{0}\rangle}\\ {L
_n}{\Phi _{h,\bar{h}}}(0,0){\vert{0}\rangle}&=& 0\nonumber\\
{\bar{L} _m}{\Phi _{h,\bar{h}}}(0,0){\vert{0}\rangle}&=&
0\nonumber
\end{eqnarray}
where $L _n$ and $\bar{L} _m$ are respectively the usual left and
right Virasoro modes and $\Phi_{h,\bar{h}}(z,\bar z)$ is a primary
conformal field representation of conformal scale $h$+ $\bar{h}$
and conformal spin $h$- $\bar{h}$. The primary $so(1,2)$ highest
weight states $\vert{s}\rangle$ and $\vert{\bar{s}}\rangle$
eqs(4.3-4) are respectively in one to one correspondance with the
left Virasoro primary state $\Phi _{h}(0){\vert{0}\rangle}
=\vert{h}\rangle$ and the right Virasoro primary one ${\Phi
_{\bar{h}}}(0){\vert{0}\rangle}=\vert {\bar{h}}\rangle$. On the
other hand, if we respectively associate to HWR(I) and HWR(II) the
mode operators $Q_{s+n}^{+}= Q_{s+n}$  and $Q_{-s-n}^{-}=\bar{Q}
_{s+n}$ and using $ SO(1,2)$ tensor product properties, one may
build, under some assumptions, an extension $\bf S $ of the $
so(1,2)$ algebra going beyond the standard supersymmetric one. To
that purpose, note first that the system { $J_0$, $J_+$ ,$J_-$ and
$Q_{s+n}$ obey the following commutation relations $(s=-1/k)$.
\begin{eqnarray}
\lbrack {J_0,Q_{s+n}}\rbrack &=&(s+n)Q_{s+n}\nonumber\\ \lbrack
{J_+,Q_{s+n}}\rbrack &=&\sqrt {(2s+n)(n+1)}Q_{s+n+1}\\ \lbrack
{J_-,Q_{s+n}}\rbrack &=&\sqrt {(2s+n-1)n}Q_{s+n-1}\nonumber
\end{eqnarray}
Similarly we have for the antiholomorphic sector:
\begin{eqnarray}
\lbrack {{\bar{J}} _0,{\bar{Q}} _{s+n}}\rbrack &=&-(s+n){\bar{Q}}
_{s+n}\nonumber\\ \lbrack {{\bar{J}} _+,{\bar{Q}} _{s+n}}\rbrack
&=&-{\sqrt {(2s+n-1)n}}{{\bar{Q}} _{s+n-1}}\\ \lbrack {{\bar{J}}
_{-},{\bar{Q}} _{s+n}}\rbrack &=&-{\sqrt {(2s+n)(n+1)}}{{\bar{Q}}
_{s+n+1}}\nonumber
\end{eqnarray}
To close these commutations relations with the $Q_s$'s through a
k-th order product one should fullfil the following constraints.
\\(1) the generalized algebra $\bf S$ we are looking for should be
a generalisation of what is known in two dimensions, i.e a
generalisation of 2d fsusy.\\(2) When the charge operator
$Q_{s+n}$ goes arround an other, say $Q_{s+m}$, one picks a phase
$\Phi ={2i{\pi} /k}$; i.e:
\begin{equation}
Q_{s+n}Q_{s+m}=e^{\pm 2i{\pi} s}Q_{s+m}Q_{s+n}+ \ldots;\quad  s=
{-{1\over k}},
\end{equation}
where the dots refer for possible extra charge operators of total
$J_0$ eigenvalue (2s+n+m). Eq (4.11) shows also that the algebra
we are searching has a ${\bf Z}_k$ graduation. Under this discrete
symmetry, $Q_{s+n}$ carries a $+1 (mod k)$ charge while the
$P_{0,\pm}$ energy momentum components have a zero charge mod k.
\\(3) the generalised algebra $\bf S $ should split into a bosonic
B part and an anyonic A and may be written as: ${\bf S} =
{\oplus_{r=0}^{k-1}} A_r  = B {\oplus_{r=1}^{k-1}} A_r $ . Since
$A_{n}A_m \subset A_{(n+m) (modk)}$ one has:
\begin{eqnarray}
{\lbrace {A_r}{\ldots} {A_r} \rbrace}_{k}&{\subset}&
{B}\nonumber\\ \lbrack {B,A}\rbrack &{\subset}& {A}\\ \lbrack
{B,B}\rbrack &{\subset}& {B}\nonumber.
\end{eqnarray}
In these eqs, ${\lbrace {A_r}{\ldots} {A_r} \rbrace}_{k}$ means
the complete symmetrisation of the k anyonic operators $A_r$ and
is defined as:
\begin{equation}
{\lbrace {A_{s_r}}{\ldots} {A_{s_r}\rbrace}_{k} ={1\over k!}{\sum
_{\sigma {\in} {\Sigma}}}{(A_{s_{\sigma (1)}}{\ldots}
{A_{s_{\sigma(k)}}}}})
\end{equation}
where the sum is carried over the k elements of the permutation
group  $\lbrace {1, {\ldots},k}\rbrace$. \\(4) the algebra $\bf S$
should obey generalised Jacobi identities; in particular one
should have:
\begin{equation}
adB{\lbrace {A_{s_1}}{\ldots} {A_{s_k}}\rbrace}=0,
\end{equation}
where B stands for the bosonic generators $J_{0,{\pm}}$ or
$P_{0,{\pm}}$ of the Poincar\'{e} algebra. Using eq(4.12) to write
${\lbrace {A_r} {\ldots} {A_r}\rbrace}_k$ as $\alpha
_{\mu}P^{\mu}+ \beta _{\mu}J^{\mu}$ where $\alpha$ and $\beta$ are
real constants; then putting back into the above relation we find
that ${\lbrace {A_r} {\ldots} {A_r}\rbrace}_k$ is proportional to
$P_{\mu}$ only. In other words, $\beta _{\mu}$ should be equal to
zero; a property which is easily seen by taking $B=P_{\mu}$ in eq
(4.14). Put differently the symmetric product  of the $D_{s}^\pm$,
denoted hereafter as $S^k[D_{s}^\pm]$, contains the space time
vector representation $D_1$ of $so(1,2)$ and so the primitive
charge operators $Q_{-{1/k}}$ and $\bar{Q}_{{1/k}}$ obey:
\begin{eqnarray}
\lbrack {J_0,(Q_{-{1/k}})^k} \rbrack&=&-(Q_{-{1/k}})^k{\sim}
P_-\nonumber\\ \lbrack {J_-,(Q_{-{1/k}})^k} \rbrack&=&0
\end{eqnarray}
Similarly we have:
\begin{eqnarray}
\lbrack {\bar{J}_0,({\bar{Q}}_{{1/k}})^k }\rbrack&=&({\bar{Q}}
_{{1/k}})^k{\sim} P_+\nonumber\\ \lbrack {\bar{J} _+,({\bar{Q}}
_{1/k})^k} \rbrack& =&0
\end{eqnarray}
Moreover acting on $(Q _{-{1/k}})^k $ by $ad{J_{+}^n}$ and on
$(\bar{Q} _{{1/k}})^k$ by $ad{ \bar{J}_{+}^n}$, one obtains:
\begin{eqnarray}
ad{J_+} (Q _{-{1/k}})^k&{\sim}& P_0\nonumber\\
 {ad{\bar{J}} _-} ({\bar{Q}} _{{1/k}})^k&{\sim}& P_0\\
 {{ad}^2 {J_+}} (Q _{-{1/k}})^k&{\sim}& {P_-}\nonumber\\
 {{ad}^2{\bar{J}} _-} ({\bar{Q}} _{{1/k}})^k&{\sim}&{P_+}\nonumber.
\end{eqnarray}
In summary, starting from $P_{(1,2)}$ and the two Verma modules
HWR(I) and HWR(II) (4.3-4 ), one may build the following new
extended symmetry:
\begin{eqnarray}
{\{ Q_{-{1\over k}}^{\pm}},{ Q_{-{1\over k}}^{\pm},\ldots
,Q_{-{1\over k}}^{\pm}\}}_{k}&=&P_{\mp}
=P_{1}{\pm}iP_{2}\nonumber\\ \{ Q_{-{1\over k}}^{\pm},\ldots
,Q_{-{1\over k}}^{\pm},Q_{1-{1\over k}}^{\pm}\}_{k} &=&\pm i\sqrt
{2\over k}P_{0}\nonumber\\
 - (k-1)\{ Q_{-{1\over k}}^{\pm},\ldots ,Q_{-{1\over k}}^{\pm},Q_{1-{1\over k}}^{\pm},Q_{1-{1\over k}}^{\pm}\}_{k}\nonumber\\
 \pm i\sqrt {k-2}\{ Q_{-{1\over k}}^{\pm},\ldots ,Q_{-{1\over k}}^{\pm},Q_{1-{1\over k}}^{\pm},Q_{2-{1\over k}}^{\pm}\}_{k}&=&P_{\pm}\nonumber\\
\lbrack J^{\pm},\lbrack J^{\pm},\lbrack J^{\pm},(Q_{-{1\over
k}}^{\pm})^{k}\rbrack \rbrack \rbrack &=&0
\end{eqnarray}
\begin{center}
\ldots
\end{center}

These eqs define the RdTS algebra. For more details on this
algebraic structure, see [1,23].

\subsection{More on RdTS invariance}
Having described RdTS symmetry, we turn now to make three comments
on the algebra (4.18).
\subsubsection{ Bulk HWR}
A generic highest weight representation of the RdTS symmetry is
obtained, as usual, by successive applications of the creation
operators on a given HW state $\vert {\Lambda}\rangle$. To work
out these representations in a tricky way, note the two
following:\\
 {\bf (1)} eqs(4.18) has two subalgebras $A_{(1+2){d}}^{+}$ and $A_{(1+2){d}}^{-}$ generated by (${J_{0,\pm},P_{0,\pm}, {\bf{Q}}_{-{r\over k}}^+}$, with $ 0<r<k$ ) and (${J_{0,\pm},P_{0,\pm}, {\bf{Q}}_{-{r\over k}}^-}$, with  $0<r<k$ ) respectively.Thus starting from the vaccum states $\vert {\Lambda^{\pm}_{s}}\rangle$ in the $(\pm{s})$ spin representations of so(1,2), one can build the on shell HW representations of $A_{(1+2){d}}^{\pm}$ which read as:\\
{\bf (i)} HWR of $A_{(1+2){d}}^+$\\
\begin{equation}
\begin{array}{rcl}
 {\bf{Q}}^{-}_{-{1\over k}}\vert \Omega^{+}_{s}\rangle &=& 0\\
\\
 \{{({\bf{Q}}^{+}_{-{1\over k}})^{r}\over [r]!}\vert \Omega^{+}_{s}\rangle &;& o\le r\le k-1\}
\end{array}
\end{equation}
Note that [r] is the q-number given by $
{{\omega^{-r}}-{\omega^r}}\over {{\omega^{-1}}-{\omega^1}}$, with
$\omega= exp{i{\pi\over k}}$. Note also that the above basis
states is k dimensional and carries the spin values $(s-{r\over
k})$.\\ {\bf (ii)} HWR of $A_{(1+2){d}}^-$\\
\begin{equation}
\begin{array}{rcl}
{ \bf{Q}}^{+}_{-{1\over k}}\vert \Omega^{-}_{-s}\rangle &=& 0\\
\\
\{{({\bf{Q}}^{-}_{-{1\over k}})^{r}\over [r]!}\vert
\Omega^{-}_{-s}\rangle &;& o\le r\le k-1\}
\end{array}
\end{equation}
In this case the basis states carry opposite values of the spin;
i.e $(-s+{r\over k})$.\\ $\bf (2)$ A class of HW representations
of the full algebra may be obtained from the above ones just by
requiring CPT invariance; in particular by imposing,\\
\begin{equation}
\begin{array}{rcl}
 {\bf{Q}}^{-}_{-{1\over k}}&=&[ {\bf{Q}}^{+}_{-{1\over k}}]^+ \\
\\
  {\bf{Q}}^{-}_{1-{1\over k}}&=&[ {\bf{Q}}^{+}_{1-{1\over k}}]^+
\end{array}
\end{equation}
In the rest frame where $ P_{0}=E$ is a constant and
$J_{\pm}={P_{\pm}}=0$, HW representations of $A_{(1+2)d}^{+}$ have
k dimensions and depend on the value of $E$. For $E=0$, the
representations are nilpotent since $( {\bf{Q}}_{-1/k}^{+})^k=0$.
However for $E$ non zero, $A_{(1+2)d}^{+}$ HWRs we are interested
in are still nilpotent but have a non trivial center. Examples of
$A_{(1+2)d}^{+}$ representations have been already studied in [1]
using finite dimensional matrix representations and differential
operators on the space of functions on $R^{1,2}$. A simple example
of matrix representation is given by,
\begin{equation}
\begin{array}{rcl}
 {\bf{Q}}_{-{1\over k}}^{+}&=&\left(\matrix{0&0&0&\ldots&0\cr \sqrt{[1]}&0&0&\ldots&0\cr0&\sqrt{[2]}&0&\ldots&0\cr\vdots&\vdots&\ddots&\ddots&\vdots\cr0&0&\ldots&\sqrt{[k-1]}&0\cr}\right),\\
\\
  {\bf{Q}}_{1-{1\over k}}^{+}&=&\left(\matrix{0&0&0&\ldots&\{\sqrt{[k-1]!}\}^{-1}\cr 0&0&0&\ldots&0\cr0&0&0&\ldots&0\cr\vdots&\vdots&\ddots&\ddots&\vdots\cr 0&0&\ldots&0&0\cr}\right)
\end{array}
\end{equation}
\subsubsection{ Edge HWR}
If one supposes that the RdTS algebra given above is also valid
for some class of (1+2) dimensional manifold $M$ with a non zero
boundary $\partial{M}$, then one may also define the limit of
eqs(4.18) on the border of $M$. To derive this limit and its
representations on $\partial{M}$, one has to first specify the
constraints giving the limit $\partial{M}$, identify its right
variables; then replace the generators of the bulk symmetry (4.18)
by the appropriate variables on $\partial{M}$. This operation
depends however on the nature of the manifold $M$ if one wants to
write down explicit formula. Nevertheless if we admit that on the
border of M, we have a two dimensional Poincar\'e invariance with
$SO(2)$ as a Lorentz subgroup; say the little group of $ SO(1,2)$
for example, we may obtain a limit of the RdTS algebra on
$\partial{M}$ just by requiring that on the border of $M$, one is
allowed to naively set $ P_{0}={J_{\pm}}=0$ in eq(4.18). Roughly
speaking the limiting RdTS algebra on $\partial{M}$ could be then
defined as:
\begin{equation}
\begin{array}{rcl}
\{ {\bf q}_{-{1\over k}}^{\pm},\ldots ,{\bf q}_{-{1\over
k}}^{\pm}\}_{k}&=&P_{\mp}\\
\\
\lbrack {J_0,({\bf q}_{-{1/k}})^k} \rbrack&=&-({\bf q}_{-{1/k}})^k
\\
\\
\lbrack {\bar{J}_0,({\bar{\bf q}}_{{1/k}})^k
}\rbrack&=&+({\bar{{\bf q}}} _{{1/k}})^k\\
\\
\lbrack {J_0},P_{\pm} \rbrack&=&0\\
\\
\lbrack {\bar{J}_0},P_{\pm} \rbrack&=&0\\
\\
\lbrack {{\bf q}_{-{1/k}}^{\pm}},P_{\pm} \rbrack&=&0
\end{array}
\end{equation}
Like in the bulk of $M$, the above algebra admits two subalgebras
$A_{2d}^{+}$ and $A_{2d}^{-}$ respectively generated by
$({J_{0},P_{\pm},{\bf q}_{-{r\over k}}^+}$,with $ 0<r<k )$ and
$({J_{0},P_{\pm},{\bf q}_{-{r\over k}}^-}$,with  $0<r<k )$. HW
representations of the algebra (4.23), to which we have refered to
as RdTS edge representations, are very special since they usually
factor into left and right terms. Moreover EHWS of the RdTS
algebra may be obtained from BHWR by going to $\partial{M}$ and
taking the appropriate limits. These representations are quite
similar to  those used in the study of 2d fsusy [3-7]. A system of
vectors basis of such representations may naively written down by
considering successive applications by ${\bf q}_{-{1\over k}}$ (
resp ${\bar{\bf q}}_{-{1\over k}}$) on a highest weight state
$\lambda$. The resulting multiplet of vector basis is then,
\begin{equation}
\lbrack({{\bf q}_{-1/k}^{\pm}})^r
\vert\lambda\rangle,0<r<(k+1)\rbrack
\end{equation}
Similar quantities may also be written down for the right sector
and then for the full algebra by imposing CPT invariance.
\subsubsection{Case where $\bf M$ is $\bf AdS_3$}
The consideration of the space time manifold $M$ as $AdS_3$ is due
to the fact that the $AdS_3 $ geometry has many relevant features
for the study of RdTS invariance. We propose to review hereafter
some of its useful properties:\\ {\bf (i)} In the euclidean
representation, $AdS_3$ has an $SO(1,3)$ isometry group containing
naturally the $SO(1,2)$ Lorentz symmetry of the (1+2)dimensional
$R^{1,2}$ space time.  \\ {\bf (ii)} $AdS_3$has a boundary space
which may be thought of as the real two-sphere. As one knows, it
lives on the two-sphere   boundary space time conformal field
theories. \\ {\bf (iii)} The boundary invariance on
$\partial{AdS_3}$ has $so(1,2)$ projective subsymmetries which we
will relate to the $so(1,2)$ Lorentz subgroup of the RdTS
Symmetry.\\ Using these features we have shown in [8] that the two
so(1,2) modules HWR(I) and HWR(II), considered in the building of
RdTS supersymmetry, are just special representations of the
$AdS_3$ boundary CFT. To see this relationship, let us review
briefly some elements of $ AdS_3$ geometry.\\ {\bf a. $\bf AdS_3$
manifold}\\
 The $ AdS_3$ space is given by the hyperbolic coset manifold $Sl(2,C)/SU(2)$ which may be thought of as the three dimensional hypersurface ${H_3}^+$
\begin{equation}
-{X_0}^2+{X_1}^2+{X_2}^2+{X_3}^2= -l^2
\end{equation}
embedded  in flat $ R^{1,3}$ with local coordinates { $X^0$,
$X^1$, $X^2$, $X^3$ }. This hypersurface describes a space with a
constant negative curvature ($-{1\over {l^2}}$). The parameter l
is choosen to be quantized in terms of the $ {l_s }$ fundamental
string lenght units; i.e, $l={l_s }\times k$ where k is an integer
to be interpreted later on as the Kac Moody level of the $
{so_{k}(1,2)}$ affine symmetry. To study the field theory on the
boundary of $AdS_3$, it is convenient to introduce the following
set of local coordinates of $AdS_3$:
\begin{eqnarray}
\phi&=&{log(X_{0}+X_{3})}/ {l}\nonumber\\ {\gamma} &=&{{X_2+iX_0}
\over {X_0+iX_3}} \\ {\bar{\gamma}} &=&{{X_2-iX_1}\over
{X_{0}+iX_3}}\nonumber
\end{eqnarray}
An equivalent description of the hypersurface is:
\begin{eqnarray}
{\gamma}&=&{r\over {\sqrt{l^2+r^2}}} e^{-\tau +i\theta
}\nonumber\\ {\bar{\gamma}}&=&{r \over {\sqrt{l^2+r^2}}}
e^{-\tau-i\theta}\nonumber\\ {\phi}&=&{\tau} +1/2log(1+r^2/l^2)\\
r&=&{le^{\phi}}\sqrt{\gamma \bar{\gamma}}\nonumber\\
\tau&=&\phi-1/2log(1+e^{2\phi}\gamma \bar{\gamma})\nonumber\\
{\theta} &=&{1\over {2i}} {log(\gamma /\bar{\gamma})}\nonumber,
\end{eqnarray}
where we have used the change of variables:
\begin{eqnarray}
X_0&=&X_0(r,\tau)={\sqrt{l^2+r^2}}cosh{\tau}\nonumber\\
X_3&=&X_3(r,\tau)={\sqrt{l^2+r^2}}sinh{\tau}\\
X_1&=&X_1(r,\theta)=rsin{\theta}\nonumber\\
X_2&=&X_2(r,\theta)=rcos{\theta}\nonumber
\end{eqnarray}
In the coordinates $( \phi, \gamma, \bar\gamma )$, the metric of
$H_3^+$ reads as:

\begin{equation}
ds^2=k(d{\Phi}^2+e^{2{\Phi}}{d{\gamma}}{d{\bar{\gamma}}})
\end{equation}
 Note that in the $( \phi, \gamma, \bar\gamma )$ frame, the boundary of euclidean $AdS_3$ corresponds to take the field $\Phi $ to infinity. As shown on the above eqs,this is a two sphere which is locally isomorphic to the  complex plane parametrized by $ (\gamma,\bar{\gamma})$ .\\
{\bf b. Strings on $\bf AdS_3$}\\ Quantum field theory on the $
AdS_3$ space is very special and has very remarkable features
governed by the Maldacena correspondence in the zero slope limit
of string theory[31 ]. On this space it has been shown that bulk
correlations functions of quantum fields find natural
interpretations in the conformal field theory on the boundary of
$AdS_3$[32]. In algebraic language, this correspondance transforms
world sheet symmetries of strings on $AdS_3$ into space time
infinite dimensional invariances on the boundary of $AdS_3$. In
what follows we give some relevant results. \\ {\bf b.1
General}\\
 To work out explicit field theoretical realisations of these symmetries, we start by recalling that in the presence of the Neveu-Schwarz $B_{\mu \nu}$ field with euclidean world sheet parameterized $(z,\bar{z})$, the dynamics of the bosonic string on $AdS_3$ is described by the following classical lagrangian:

\begin{equation}
 L=k[{\partial {\Phi}}{\bar{\partial}}{\Phi}+e^{2{\Phi}}{\partial {\gamma}}{\partial {\bar{\gamma}}}]
\end{equation}
 In this eq ${\partial}$ and $\bar{\partial}$ stand for derivatives with respect to z and $\bar{z}$ repectively. Introducing two auxiliary variables ${\beta}$ and ${\bar{\beta}}$, the above eq may be put into the following  convenient form:

\begin{equation}
 L^{'}=k^{2}({\partial {\Phi}}{\bar{\partial} {\Phi}}+{\beta {\bar{\partial}} \gamma}+{\bar{\beta}}{\partial {\bar{\gamma}}}-e^{-2{\Phi}}
{\beta {\bar{\beta}}})
\end{equation}
The eqs of motion of the various fields one gets from eq(4.31)
read as:

\begin{eqnarray}
{{\partial}{\bar{\partial}}}{\Phi} -2{\beta
{\bar{\beta}}}e^{-2{\Phi}}&=&0\nonumber\\
{\bar{\partial}}\gamma-\beta e^{-2{\Phi}}&=&0\\
\partial{\bar{\gamma}}-{\bar{\beta}} e^{-2{\Phi}}&=&0\nonumber\\
\partial{\bar{\beta}}&=&{\bar{\partial}}\beta=0\nonumber
\end{eqnarray}
 String dynamics on the boundary of $AdS_3$ is obtained from the previous bulk eqs by taking the limit $\Phi$ goes to infinity. This gives:

\begin{eqnarray}
{{\partial}{\bar{\partial}}}\Phi &=&0\nonumber\\
{\bar{\partial}}\gamma&=&\partial {\bar{\gamma}}=0\\
\partial{\bar{\beta}}&=&\bar{\partial}\beta =0\nonumber
\end{eqnarray}
The world sheet (WS) fields $\Phi, \gamma$ and $\bar{\gamma}$
which had general expressions in the bulk become now holomorphic
on the boundary of $AdS_3$ and describe a boundaryCFT. Note that
consistency of quantum mechanics of the string propagating in
space time requires that the target space should be
$AdS_3{\times}N$, where N is a (3+n) dimensional compact manifold.
To fix the ideas, N may be thought of as $S^3{\times}T^n$ with $n
=20$ for the bosonic string and $n= 4$ for superstrings.\\ {\bf
b.2  WS Symmetries}\\
 World sheet invariances include affine Kac-Moody, Virasoro symmetries and their extensions. For a bosonic string propagating on $AdS_3{\times}S^3{\times}T^{20}$, we have the following:\\

${\bf \alpha}$.  Three kinds of WS affine Kac-Moody invariances:\\

$(\bf i)$ A level $(k-2)$ ${sl(2)}\times{\bar{sl(2)}}$ invariance
coming from the string propagation on $AdS_3$. This invariance is
generated by the conserved currents $J_{sl(2)}^q$ and
${{\bar{J}}_{sl(2)}}^{q}; q=0,{\pm} 1$. In terms of the WS fields
${\Phi, \gamma, \bar{\gamma},\beta} $ and $\bar{\beta}$ of
eq(4.32), the field theoretical realization of these currents is
given by the Wakimoto representation:

\begin{eqnarray}
{J^{-}}(z)&=&{\beta}(z)\nonumber\\
J^{+}(z)&=&{\beta}{{\gamma}^{2}}+{\sqrt{2(k-2)}}{\gamma}{\partial
{\Phi}} + k{\partial {\gamma}}\nonumber\\
J^{0}(z)&=&{\beta}{\gamma}+1/2{\sqrt{2(k-2)}}{\partial {\Phi}}\\
{\bar{J}}^{-}(\bar{z})&=&\bar{\beta}\nonumber\\
{\bar{J}^0}(\bar{z})&=&{\bar{\beta}}{\bar{\gamma}}+1/2{\sqrt{2(k-2)}}
{\partial {\Phi}}\nonumber \\
{{\bar{J}}^{+}}(\bar{z})&=&{\bar{\beta}}{\bar{\gamma}}^{2}+{\sqrt{2(k-2)}}{\bar{\gamma}}{\partial
{\Phi}} + k{\partial {\bar{\gamma}}}\nonumber
\end{eqnarray}
$(\bf ii)$ A level $(k+2)$ invariance coming from the string
propagation on $S^3$. The conserved currents are $J_{su(2)}^q$ and
${\bar{J}}_{su(2)}^q$. The WS field theoretical realization of
these currents is given by the level $(k+2)$  WZW $su(2)$ model
[33 ].\\ $(\bf iii)$ A $u(1)^{20}{\times}{\bar{u}(1)}^{20}$
invariance coming from the torus $T^{20}$. This symmetry is
generated by $20$ $ U(1)$ Kac Moody currents
$J_{u(1)}^i;i=1,\ldots,20$.\\

${\bf \beta}$.  WS Virasoro symmetry\\

This symmetry, which splits into holomorphic and antiholomorphic
sectors, is given by the Suggawara construction using quadratic
Casimirs of the previous WS affine Kac Moody algebras. For the
holomorphic sector, the WS Virasoro currents of a bosonic string
on ${AdS_3}\times {S^3}\times {T^{20}}$ are:\\ $(\bf i)$   String
on $ AdS_3 $:

\begin{equation}
T_{sl(2)}^{WS}={1\over
(k-2)}[(J_{sl(2)}^0)^{2}-(J_{sl(2)}^{1})^{2}-(J_{sl(2)}^{2})^{2}]
\end{equation}
$(\bf ii) $  String on $ S^3 $:
\begin{equation}
T_{su(2)}^{WS}={1\over
(k+2)}[(J_{su(2)}^0)^{2}+(J_{su(2)}^{1})^{2}+(J_{su(2)}^{2}) ^{2}]
\end{equation}
$(\bf iii)$   String on $ T^{20} $:
\begin{equation}
T_{u(1)}^{WS}={\sum_{i=1}^{20}}{[J_{u(1)}^i]}^2
\end{equation}

Similar quantities are also valid for the antiholomorphic sector
of the conformal invariance. Note that the total WS energy
momentum tensor $T_{tot}^{WS}$ is given by the sum of
$T_{sl(2)}^{WS}, T_{su(2)}^{WS}$ and $T_{u(1)}^{WS}$.\\ In the
case of a superstring propagating on
$AdS_3{\times}S^3{\times}T^4$, the above conserved currents are
slightly modified by the adjunction of extra terms due to
contributions of WS fermions .
\section{Infinite Space-time invariances of $\bf AdS_3$}
To analyse the space-time infinite dimensional symmetries on the
boundary of $AdS_3$, one may follow the same strategy that we have
used for the study of WS invariances. First identify the space
time affine Kac-Moody symmetries and then consider the space time
conformal invariance and eventually the Casimirs of higher ranks.
We shall simplify a little bit the analysis of space-time
invariance and focus our attention on the conformal symmetry on
$\partial ({AdS_3})$. Some specific features on space time
Kac-Moody symmetries will also be given in due time.\\ We begin by
noting that space time infinite invariances on the boundary of
$AdS_3$ are intimately linked to the WS ones. For the case of a
bosonic string propagating on $AdS_3{\times}S^3{\times}T^{20}$, we
have already shown that there are various kinds of WS symmetries
coming from the propagation on $AdS_3 , S^3$ and $T^{20}$
respectively. In the $\phi$ infinite limit, we can show that one
may use these WS symmetries to build new space time ones.
 \subsection { Space time conformal invariance}
First of all, note that the global part of the WS
$SO(1,2){\times}{\bar{SO(1,2)}}$ affine invariance of a bosonic
string on $AdS_3$, generated by $J_{0}^q$ and ${\bar{J}}_{0}^q;
q=0,\pm 1$  may be  realized in different ways. A tricky way,
which turns out to be crucial in building space-time conformal
invariance, is given by the Wakimoto realization [34].
Classically, this representation reads in terms of the local
coordinates $(\Phi, \gamma, \bar{\gamma})$ as follows:
\begin{equation}
\begin{array}{lcr}
J_{0}^0=\gamma {{\partial}/{\partial{\gamma}}}-
1/2{{\partial}/{\partial{\gamma}}},\\
J_{0}^-={\partial}/{\partial{\gamma}},\\ J_{0}^+={\gamma}^{2}
{{\partial}/{\partial{\gamma}}}-{\gamma}
{{\partial}/{\partial{\Phi}}}-e^{-2{\Phi}}{\partial}/{\partial{\gamma}}.
\end{array}
\end{equation}
Similar relations are also valid for $\bar{J}_{0}^q$; they are
obtained by substituting $\gamma$ by $\bar{\gamma}$. Quantum
mechanically, the charge operators $J_{0}^q$ and $\bar{J}_{0}^q$
are given in terms of the Laurent mode operators of the quantum
fields $\Phi, \gamma, \bar{\gamma}, \beta$ and $\bar{\beta}$ by
using eqs(4.36) and performing the Cauchy integrations:
\begin{equation}
\begin{array}{lcr}
J_{0}^q=\int {dz\over 2i\pi}J^q(z)\\ \bar{J}_{0}^q=\int
{d{\bar{z}} \over {2i\pi}}{\bar{J}}^{q}(\bar z).
\end{array}
\end{equation}
To build the space time conformal invariance on the $AdS_3$
boundary, we proceed by steps. First suppose that there exists
really a conformal symmetry on the boundary of $AdS_3$ and denote
the space time Virasoro generators by $L_{n}$ and  $\bar{L}_{n},
n{\in} Z$. The ${L}_{n}$ and  ${\bar{L}}_{n}$, which should not be
confused with the WS conformal mode generators, satisfy obviously
the left and right Virasoro algebras.
\begin{equation}
\begin {array}{lcr}
[L_{n}, L_{m}]={(n-m)}L_{n+m}+{c\over 12}n{(n^{2}-1)} {\delta
_{n+m}}\\

[{\bar{L}}_{n},{\bar{L}}_{m}]=(n-m)\bar{L}_{n+m}+{{\bar{c}}\over
12}n(n^{2}-1) {\delta_{n+m}}\\

[L_{n},{\bar{L}}_{m}]=0 .
\end {array}
\end{equation}
The second step is to solve these eqs by using the string WS
fields $(\Phi, \gamma, \bar{\gamma})$  on $AdS_3$. To do so, it is
convenient to divide the above eqs into two blocks. The first
block corresponds to set $ n=0,\pm 1$ in the generators $L_n$ and
$\bar L_n$ of eqs(6.3). It describes the anomaly free projective
subsymmetry the Virasoro algebra. The second block concerns the
generators associated with the remaining values of n.\\ On the
boundary of $ AdS_3$ obtained by taking the infinite limit of the
$\Phi$  field, one solves the projective subsymmetry by natural
identification of $L_{q}$ and ${\bar{L}}_q$ ; $q=0,\pm 1$ with the
zero modes of the WS $so(1,2){\times}{\bar{so}}(1,2)$ affine
invariance. In other words we have:
\begin{equation}
\begin{array}{lcr}
{L}_q&=&-\int {dz\over 2i\pi}J^q(z) =-J_{0}^q;\quad q=0,\pm 1\\
{\bar{L}}_q&=&\int {d{\bar{z}}\over 2i\pi}\bar{J}^q(\bar z)
=-{\bar{J}}_{0}^q; \quad  q=0,\pm 1.
\end{array}
\end{equation}
Note that on the $AdS_3$ boundary, viewed as a complex plane
parametrized by $( \gamma, \bar{\gamma} )$, the charge operators
$J_{0}^-$ ( $L _{-1}$ ) and $\bar{J}_{0}^{-}(\bar{L}_{-1})$ taken
in the Wakimoto representation  coincide respectively with the
translation operators $P_{-}$ and $\bar{P}_{+}$:

\begin{equation}
\begin{array}{lcr}
P _{-}=L_{-}= {{\partial}/{\partial {\gamma}}}\\ P
_{+}={\bar{L}}_{-}= {{\partial}/{\partial {\bar{\gamma}}}}.
\end{array}
\end{equation}
Eqs(5.4-5) are interesting; they establish a link between the
$L_{-}$ and ${\bar{L}}_{-}$ constants of motion of the boundary
conformal field theory on $AdS_{3} $ on one hand and the $P _{-}$
and the $P _{+}$ translation generators of the RdTS extension of
the $so(1,2)$ algebra on the other hand. \\ To get the rigourous
solution of the remaining Virasoro charge operators ${L}_{n}$ and
${\bar{L}}_{n}$, one has to work hard. This is a lengthy and
technical calculation which has been done in [35] in connection
with the study of the $D_{1}/D_{5}$ brane system. We shall use an
economic path to work out the solution. This is a less rigourous
but tricky way to get the same result. This method is based on
trying to extend the $L_{n}$ and $ \bar{L}_{n};\quad  n=0,{\pm 1}$
projective solution to arbitrary integers n using properties of
the string WS fields near the boundary, dimensional arguments and
similarities with the photon vertex operator in three dimensions.
Indeed using the holomorphic property of $\gamma $ and
$\bar{\gamma}$ eqs(4.33) as well as the space time dimensional
arguments;
\begin{equation}
\begin{array}{lcr}
 \lbrack \gamma \rbrack = -1;\quad  J_{sl(2)}^{0} = 0\\
 J_{sl(2)}^{-}= 1 ;\quad  J_{sl(2)}^{+}= -1,
\end{array}
\end{equation}
it is not difficult to check that the following $L_{n}(\bar{L}
_{n})$ expressions are good condidates:

\begin{equation}
{\bf L}_{n}=\int {dz\over {2i\pi}}\lbrack
a_{0}{{\gamma}^{n}}J_{sl(2)}^{0}- { a_{-}\over 2}
{{\gamma}^{n+1}}J_{sl(2)}^{-}+ { a_{+}\over 2}
{{\gamma}^{n-1}}J_{sl(2)}^{+}\rbrack,
\end{equation}
and a similar relation for $\bar{L} _n$. To get the $a_i$
coefficients, one needs to impose constraints which may be
obtained by using results of BRST analysis in $QED_3$. Following
[32],the right constraints one has to impose on the $ a_i $' s
are:

\begin{equation}
\begin{array}{lcr}
    n a_{0}  + (n+1 ) a_{-} + (n-1 ) a_{+}= 0\\
    J^{0}{\gamma}-(1/2) J^{-}{\gamma^2}-(1/2) J^{+} = 0.
\end{array}
\end{equation}
The solution of the first constraint of these eqs reproducing the
projective generators (5.4) is as follows:

\begin{equation}
\begin{array}{lcr}
 a_{0}= ({n^2}-1)\\
a_{-}= n(n-1)\\ a_{+}= n(n+1)
\end {array}
\end{equation}
Moreover using the second constraint of eqs(5.8) to express
$J_{sl(2)}^{+}(z)$ in terms of $J_{sl(2)}^{0}(z)$ and
$J_{sl(2)}^{-}(z)$; then putting back into eqs(5.7), we find:

\begin{equation}
{\bf L}_{n}=\int {dz\over {2i\pi}}\lbrack {-(
n+1){{\gamma}^{n}}J_{sl(2)}^{0}+ n
{{\gamma}^{n+1}}J_{sl(2)}^{-}}\rbrack .
\end{equation}
Eqs (5.4) and (5.10) define the space time Virasoro algebra on the
boundary of $AdS_3$.
\subsection{ Other symmetries of $\bf AdS_3$}
Having built the $L_n$'s space time Virasoro generators, one may
be interested in determining the space-time energy momentum
tensors $T({\gamma})$ and $\bar{T}{({\bar\gamma})}$ of the
boundary CFT on $AdS_3$. It turns out that the right form of the
space-time energy momentum tensor depends moreover on  auxiliary
complex variables $(y,\bar{y})$ so that the space time energy
momentum tensor has now two arguments; i.e: $T=T(y,\gamma)$ and
$\bar{T}= \bar{T}({\bar{y}},\bar{\gamma})$. Following [33],
$T(y,\gamma )$ and $\bar{T}({\bar{y}},\bar{\gamma})$ read as:

\begin{equation}
\begin{array}{lcr}
T(y,\gamma)= \int {dz\over{2i\pi}}\lbrack
{{\partial_{y}{J(y,\gamma)}}\over{{(y-\gamma)}^{2}}}
-{{{{\partial^{2}}_{y}}J(y,\gamma)}\over{(y-{\gamma})}}\rbrack\\
\bar{T}({\bar{y}},{\bar\gamma})=\int
{d{\bar{z}}\over{2i\pi}}\lbrack
{\partial_{\bar{y}{J(\bar{y},\bar{\gamma})}}\over{(\bar{y}-\bar{\gamma})}^{2}}-{{{\partial^{2}}_{\bar{y}}{J(\bar{y},\bar{\gamma})}}\over{(\bar{y}-\bar{\gamma})}}
\rbrack,
\end {array}
\end{equation}
where the currents $J(y,\gamma)$ and $J(\bar{y},\bar\gamma)$ are
given by:
\begin{equation}
\begin {array}{lcr}
J(y,\gamma
)=-J^{+}(y,\gamma)=2yJ^{0}(\gamma)-J^{+}(\gamma)-y^{2}J^{-}(\gamma).
\end{array}
\end{equation}
In connection to these eqs, it is interesting to note that the
conserved currents $J^{q}(y,\gamma)$ and
$J^{q}(\bar{y},\bar{\gamma})$ are related to the WS affine
Kac-Moody ones on $AdS_3$ as follows:

\begin{equation}
\begin{array}{rcl}
J^{+}(y,\gamma)&=& e^{-yJ_{0}^{-}}J^{+}(\gamma)e^{yJ_{0}^{-}}\\
&=&J^{+}(\gamma)-2yJ^{0}(\gamma)+y^{2}J^{-}(\gamma)\\
J^{0}(y,\gamma)&=&e^{-yJ_{0}^{-}}J^{0}(\gamma)e^{yJ_{0}^{-}}\\
&=&J^{0}(\gamma)-yJ^{-}(\gamma)={-{1\over 2}}{\partial
_z}J^{+}(y,\gamma)\\
J^{-}(y,\gamma)&=&e^{-yJ_{0}^{-}}J^{-}(\gamma)e^{yJ_{0}^{-}}\\
&=&J^{-}(\gamma)={1\over 2}{{\partial }^{2} _z}J^{+}(y,\gamma)
\end {array}
\end{equation}
and analogous eqs for $J^{q}(\bar{y},\bar{\gamma})$. Putting
eqs(5.12) back into eqs(5.11) and expanding in power series of
$\gamma \over{y}$, one discovers the $ L_n$ space time Virasoro
generators given by eqs( 5.10).\\ Note moreover that one may also
build space time affine Kac-Moody symmeties out of the WS ones.
Starting from WS conserved currents $E_{ws}^a(z)$, which may be
thought of as $ J_{sl(2)}^{q}(z)$, and going to the boundary of
$AdS_3$, the corresponding space time affine Kac-Moody currents
$E_{space time}^a (y,\gamma)$ read as:

\begin{equation}
E_{space time}^a(y,\gamma)= \oint  {dz\over 2i\pi}\lbrack
{{E_{ws}^{a}(z)}\over{(y-\gamma(z))}}\rbrack.
\end{equation}
Expanding this eq in powers of $y\over {\gamma}$ or $\gamma \over
{y}$, one gets the space time affine Kac-Moody modes:

\begin{equation}
E_{n}^{a,space time}= \oint {dz\over {2i\pi}}\lbrack
E_{ws}^{a}(z){\gamma}^{n}\rbrack.
\end{equation}
Note finally that RdTS symmetry may be thought of as just integral
deformations of WS and space time symmeties on $AdS_3$. Such
procedure is standard in the study of integrable models obtained
from deformations of CFTs by relevant perturbations. To see how
this works in our case let us give some details. If one forgets
about string dynamics as well as the nature of the compact
manifold $ N$ and just retains that on $\partial({AdS_{3}})$ lives
a conformal structure, one may consider its highest weight
representations which read as in eqs(3.23). A priori the central
charge c and the conformal weights h and $\bar{h}$ of these
representations are arbitrary. However requiring unitary
conditions, the parameters c, h and $\bar{h}$ are subject to
constraints which become more stronger if one imposes extra
symmetries such as supersymmetry or parafermionic invariance [36].
Having these details in mind, one may also build descendant states
$\vert {{h+n},{\bar{h}+\bar{n}}}\rangle$ of $\vert
{h,{\bar{h}}}\rangle$ from the primary ones as follows,
\begin{equation}
 \vert {{h+n},{\bar{h}+\bar{n}}}\rangle={\sum_ {\stackrel{n=\sum {\alpha _{i}} n_{i}}{\bar{n}=\sum {\beta _{j}} n_{j}}}\lambda _{\{\alpha_{i}\} \{\beta _{j}\} }{(\Pi _{i}L_{-n_{i}}^{\alpha {i}})(\Pi _{j}\bar{L}_{-n_{j}}^{\beta {j}}})} \vert {h,{\bar{h}}}\rangle.
\end{equation}
where the $\alpha_{i}$'s and $\beta_{j}$'s are positive integers
and $\lambda _{\alpha \beta }$ are C-numbers which we use to
denote the collective coefficients $\lambda _{\{\alpha_{i}\}
\{\beta _{j}\} } $. They satisfy the following obvious relations.

\begin{equation}
\begin {array}{lcr}
{L _0}{\vert {{h+n},{\bar{h}+\bar{n}}}\rangle}=(h+n){\vert
{{h+n},{\bar{h}+\bar{n}}}\rangle}\\

{L_{\pm}}{\vert
{{h+n},{\bar{h}+\bar{n}}}\rangle}=a_{\pm}(h,n){\vert {{h\pm
n},{\bar{h}\pm \bar{n}}}\rangle}\\

{{\bar{L}} _0}{\vert
{{h+n},{\bar{h}+\bar{n}}}\rangle}=({\bar{h}}+{\bar{n}}){\vert
{{h+n},{\bar{h}+\bar{n}}}\rangle}\\

{{\bar{L}}_{\pm}}{\vert {{h+n},{\bar{h}+\bar{n}}}\rangle}
={{\bar{a}}_{\pm}}(\bar{h},\bar{n}){\vert {h\pm n},{\bar{h}\pm
\bar{n}}\rangle},
\end {array}
\end{equation}
where $a_{\pm}(h,n)$ and $\bar{a} _{\pm}(\bar{h},{\bar{n}})$  are
normalization factors. Making an appropriate choice of the
$\lambda _{\alpha \beta }$ coefficients and taking the
$a_{\pm}(h,n)$ and $\bar{a} _{\pm}(h,\bar{n})$ coefficients as
given herebelow,
\begin{equation}
\begin {array}{lcr}
a_{-}(h,n)=\sqrt{(2h+n)(n+1)}\\ a_{+}(h,n)=\sqrt{(2h+n-1)n},
\end {array}
\end{equation}
one gets the two $so(1,2)$ modules eqs(4.9-10) used in building
RdTS supersymmetry. As a summary we should retain :\\ $\bf
(\alpha)$.  the RdTS extension of Poincar\'e invariance in (1+2)
dimensions we have been describing is a special kind of fsusy
algebra. It is a residual symmetry of a boundary space time
conformal invariance living on $\partial M $. Here $M= AdS_3$.\\
$\bf (\beta)$. the explicit analysis of this paper has been made
possible due to the particular properties of the $AdS_3$ geometry
since,\\
 $(\bf i)$- the $AdS_3$ manifold carries naturally a $so(1,2)$ affine invariance which has various realisation ways.\\
 $(\bf ii)$- the Wakimoto realisation of the $SO(1,2)$ affine symmetry which on one hand relates its zero mode to the projective symmetry of a boundary CFT on $AdS_3$ and on the other hand links the $L_{-}$ and $\bar L_{-}$ to the translation operators on $\partial ({AdS_3})$ as shown on eqs (5.4). \\
 $(\bf iii)$- the correspondance between WS and space time symmetries which plays a crucial role in analysing the various kinds of symmetries living on $\partial {(AdS_3)}$. \\
In the end we want to that the comodity in using $AdS_3$ geometry
in the above algebraic approach should be compared to the droplet
approximation of the CS effective field theory of FQH liquids.
\section{FQHS/RdTS Correspondance}
Let us start by recalling the different types of exotic quantum
states that we have encountered in the study FQHE systems with
boundaries and in the analysis of RdTS representations.
\subsection{FQH states}
In the CS effective model of FQHE, we have distinguished two kinds
of quantum states carrying fractional values of the spin and the
electric charge. These are the Bulk states and edge ones.\\ $(\bf
1)$ Bulk states are localised states carrying fractional spins
$\theta\over\pi$ as in eqs (2.13) and (3.6) and playing a crucial
role in (1+2)dimensional effective CS abelian $U(1)^k$ gauge
model, especially in the study of hierarchies. Classically these
states might be thought of as associated with Wilson lines of the
CS gauge fields. Quantum mechanically, these states should be
viewed as representation states of q-deformed algebras of creation
and annihilation operators of certain quantum fields. However a
such quantum fields formulation is still far from reach as no
consistent local quantum field model has been built yet.
Tentatives towards developping a q-quantum field operators
generating quantum states carrying fractional values of the spin
have been considered in different occasions in the past; for a
review on some field and algebraic methods, see [37]. For our
concern, we shall make,  In subsection 6.3, a hypothesis regarding
this matter by suggesting the RdTS symmetry as the algebra of
these quantum states.\\ $(\bf 2)$ Edge states are extended states
carrying also fractional values of the spin and are nicely
described, in the droplet approximation, by a boundary conformal
field theory. Edge states are  quantum states playing an important
role first because they are responsible for the quantization of
the Hall conductivity  $\sigma_{xy}$ and second for their non
trivial dynamics on the boundary.
\subsection{RdTS states}
 In the language of RdTS representations, we have also distinguished bulk representations living on $M$ and edge representations living on its border $\partial M$. \\
$(\bf 1)$ Bulk states are highest states of the HWRs of the
algebra (4.18). They are given by CPT invariant representations of
the $A_{(1+2)d}^{\pm}$,:
\begin{equation}
\begin{array}{lcr}
 {(\bf{Q^{+}}_{-{1\over k}})^{r}\over [r]!}\vert \Omega^{+}_{s}\rangle,
{\bf{(Q^{-}}_{-{1\over k}})^{r}\over [r]!}\vert
\Omega^{-}_{-s}\rangle ,\\ \bf{Q^{\mp}}_{-{1\over k}}\vert
\Omega^{\pm}_{s}\rangle =0,
 \bf{Q^{-}}_{-{1\over k}}=[\bf{Q^{+}}_{-{1\over k}}]^+\\
 \bf{Q^{-}}_{1-{1\over k}}=[\bf{Q^{+}}_{1-{1\over k}}]^+,
\end{array}
\end{equation}
where $ o\le r\le k-1$.\\ $(\bf 2)$ Edge states are highest states
of the HWRs of the algebra (4.23).They read as follows,
\begin{equation}
\lbrack({{\bf q}_{-1/k}^{\pm}})^r
\vert\lambda\rangle,0<r<(k+1)\rbrack.
\end{equation}
\subsection{Correspondence}
 We have learned in subsection 2.2 that in the droplet approximation, egde excitations of FQH droplets are described by conformal vertex operators carrying fractional values of the spin. These vertices live on the boundary of the disk geometry of the droplet. Abstraction done from the value of the conformal spin, the droplet vertices look like the tachyon vertex of string field theory; their correlations and scatterings may be then studied by using similar methods as those developed in the context of open string field theory [38]. \\
Outside the droplet approximation where the boundary conformal
invariance is not exact, one expects that some constants of motion
carrying fractional values of the spin are still conserved and are
the generators of the exotic quantum excitations one encounters in
the effective CS gauge model of FQHE. If one accepts this
reasonning, it follows then that there is a natural correpondence
between the edge excitations of FQHE and the RdTS representation
states living on $\partial M$. This conjecture is also supported
by the fact that representations on $\partial M$ for both FQH
liquids in the droplet approximation and RdTS symmetry on $AdS_3$
are associated with boundary conformal invariances on $\partial
M$. Beyond these approximations, the boundary conformal invariance
is no longuer exact and one is left with residual symmetries and
constants of motion carrying fractional values of the spin. In
other words we expect that one may have the following natural
correspondence: \\ {\bf (1)} bulk states involved in the CS
effective gauge model FQHE are associated with bulk HWR of the
RdTS algebra. Put differently the algebra (4.18) could be
conjectured as a condidate for the algebra of creation and
annhilation of quasiparticle states.\\ {\bf (2)} Edge states of
FQH systems are associated with edge HWR of the limit of RdTS
algebra on the border $\partial M$ of the (1+2) dimensional
manifold $M$.\\ {\bf (3)} droplet approximation of FQH liquids is
associated with the $AdS_3$ geometry. In other words the edge CFT
of FQH droplet is associated with space time CFT on the
$\partial(AdS_3$).
\section{Conclusion}
In this paper we have mainly studied two things:\\
 First, we have considered the Chern-Simons effective gauge model of FQHE and completed partial results on topological orders of FQH hierachical states. We have shown that, upon performing special $Gl(n,Z)$ transformations on the CS gauge fields, one may build an equivalent effective model having the same filling fraction $\nu$ and extending known results on Haldane hierarchy. More specifically, we have shown that any $k^{th}$ level Haldane state of filling fraction $\nu_H$ can be usually interpreted as a bound state of k Laughlin states of filling fractions ($\nu_j$); $j=1,\ldots,k$, where $\nu_j={1\over{m_{j}m_{j+1}}}$ and where the $m_j$ integers are solutions of the series $m_{j}=p_{j}m_{j-1}-m_{j-2}$; with $m_0=1$ and $m_1=p_1$. One of the remarkable properties of this decomposition is that the ${m_{j}m_{j+1}}$ product is usually an odd integer and so each factor $\nu_j$ describes indeed a Laughlin state.\\
Second we have reviewed the main lines of the RdTS algebra on a
(1+2)dimensional manifold $M$ with a boundary $\partial{M}$ and
studied its highest weight representations. We have shown that
RdTS algebra has generally two kinds of HWRs: bulk highest weight
representations and edge highest weight ones. They live
respectively in $M$ and on $\partial M$. Then using the features
of the FQH quasiparticles we encounter in the CS effective field
model of FQHE, we have conjectured that they are appropriate
condidates to be described by the HWRs of the RdTS algebra. We
have given several arguments supporting our hypothesis either by
using algebraic correspondence between FQH quasiparticles and RdTS
HW states or by working out geometric similarities between the
droplet approximation  and $AdS_3$ geometry.
  In the end we would like to note that CS effective gauge model of FQHE and the RdTS algebra have quite similar boundary invariances. In  droplet approximation of the FQH liquids as well as for RdTS problem on the $AdS_3$ space-time geometry, the above invariances share some general features with the Maldacena correspondence of strings propagating on Anti-de Sitter backgrounds[3]. We hope to be able to develop this issue in a future occasion and find ways for exploiting results, obtained in the context of strings on $AdS_{r}\times N^{10-r}$ in particular in the study of the $ D_1$/$D_5$ system of superstrings on $AdS_3$ times a seven dimensional compact manifold $N^7$ [39], in order to analyse further the correspondance we have proposed in section 5.\\
\\\\{\bf Acknowledgements}\\\\ One of us (I.B) would like to thank Prof
J Wess for kind hospitality at section der physik theorie, Munich
University and CNCPRST (Morocco) and DFG (Germany) for supporting
her stay in Munich under the contrat 445 Mar 113/5/0 Germany. This
research work has been also supported by the program under
 PARS Phys 27/ 372/98.

\section{\bf {References}}
\begin{enumerate}
\item[[1]]M.Rausch de Traubenberg and M.J. Slupinski,
          Mod.Phys.Lett. A12 (1997) 3051-3066.
\item[[2]]M.Rausch de Traubenberg, M. J. Slupinski
Fractional Supersymmetry and Fth-Roots of
Representations,J.Math.Phys.41(2000)4556-4571.\\ M.Rausch de
Traubenberg, Fractional supersymmetry and Lie Algebras, Lectures
given at the workshop on string theory and non commutative
geometry. To appear in the proceeding of the workshop June 16-17,
2000, Rabat.
\item[[3]]A.Leclair, C.Vafa Nucl.Phys. B401(1993)413. D.Bernard, A Leclair, Nucl.Phys. B340(1990)712; Phys.Lett B247 (1991)309; Commun.Math.Phys. 142 99.
\item[[4]]E.H.Saidi, M.B.Sedra and J.Zerouaoui, Class.Quant.Grav. 12(1995)1567-1580.
\item[[5]]E.H.Saidi, M.B.Sedra and J.Zerouaoui, Class Quant Grav. $\bf 12$ 2705. A.Kadiri, E.H Saidi, M.B Sedra and J.Zerouaoui 1994 On exotic supersymmetries of the thermal deformation of minimal models ICTP preprintIC/94/216. Perez A, Rausch de Traubenberg M and Simon P 1996 Nucl.Phys.B $\bf 482$ 325
\item[[6]]A.ElFallah,E.H Saidi,and J.Zerouaoui Phys.Lett.B 468(1999)86-95.Chakir,A.ElFallah and E.H Saidi,Mod Phys lett $\bf 38$ 2931 Class Quant.Grav.$\bf 14$(1997)20-40. A. ElFallah, E.H Saidi and R Dick Class Quant.Grav.$\bf 17$(2000)43
\item[[7]]I.Benkaddour and E.H. Saidi Class.Quantum. Grav.16 (1999)$1793-1804$.
\item[[8]]I Benkaddour, A ElRhalami and E.H Saidi "Non Trivial Extension of the (1+2)-Poincar\'e Algebra and Conformal Invariance on the Boundary of $\bf AdS_3$" hep-th/0007142
\item[[9]]M.D. Johnson and G.S Canright Phys. Rev.B49 (1994)2947
\item[[10]]R.E.Prange and S.M. Girvin, The Quantum Hall effect(Springer, New York, 1987), R.B.Laughlin, Phys.rev.Lett. 50(1983)1395.
\item[[11]]Xiao-Gang Wen "Topological orders and Edge Escitations in FQH State"; cond-mat/9506066
\item[[12]]C.L.Kane and Matthew P.A. Fisher "Impurity scattering and transport of Fractional Quantum Hall Edge States" new v.(july 20.2000);cond-mat/9409028
\item[[13]]R.E.Prange " The Quantum Hall Effect" 2d.Ed 1990 chapter (I) J.E.Moore and F.D. Haldane "Edge excitations of the $\nu ={2\over 3}$ spin-singlet QHS" cond-mat/9606156
\item[[14]]Steven M. Girvin,(Indiana University), "The Quantum Hall Effect: Novel Excitations and Broken Symmetries",cond-mat/9907002.
\item[[15]]F.D.M.Haldane, Phys.rev.lett.51,(1983)605
\item[[16]]B.I.Halperin, Phys.Rev.Lett.52, (1984)1583
\item[[17]]S.Girvin, Phys.Rev.B29,(1984)6012
\item[[18]]A.H.Macdonald, D.B.Murray, Phys.Rev.B32 (1985)2707
\item[[19]]J.K.Jain,Phys.RevB41(1991)7653
\item[[20]]X.G.Wen and Zee, Phys.RevB44 (1991)274
\item[[21]]B.Blok and X.G.Wen, Phys.Rev.B42 (1990)8133
\item[[22]]J.Frohlich and A.Zee,Nucl.Phys.B364, (1991)517
\item[[23]]Eduardo Fradkin, Lecture Note Series 82 (Field Theories of Condensed matter systems) 1991
\item[[24]]Shou Cheng Zhang , Int.J.Mod.Phys.B6 (1992)
\item[[25]]X.G.Wen,Int.J.Mod.Phys.B2,(1990)
\item[[26]]X.G.Wen,Q.Niu,Phys.Rev.B41, (1990)9377
\item[[27]]A ElRhalami and E.H Saidi,in preparation
\item[[28]]X.G. Wen Dynamics of the Edge excitations in the FQH Effects, In the Proceeding of the Fourth Trieste conference on Quantum Field Theory and Condensed matter Physics; edited by Randjbar-Daemi and Yu.Lu, World scientific (1991).
\item[[29]]D.H.Lee and X.G.Wen, Phys.Rev.lett.66, (1991) 1765
\item[[30]]J Wess and B. Zumino Nucl Phys B( Proc Suppl.)18(1990)302.\\C De Concini and V Kac Prog.Math 92.(1990)471\\Daniel Arnaudon, Vladimir Rittenberg, Quantum Chains with $U_q(SL(2))$ Symmetry and Unrestricted Representations, Phys.Lett. B306 (1993) 86-90
\item[[31]]J.Maldacena,Adv.Theor.Math.Phys.2 (1997)231.hep-th/9711200.
\item[[32]]A.Giveon,D.Kutasov and N.Seiberg, Comments on String Theory on $AdS_3$, hep-th/9806194
\item[[33]]J.Balog,L.O'Raifeartaigh,P. Forgacs,A.Wipf, Nucl.Phys.B325 (1989)225
\item[[34]]M.Wakimoto, Comm.Math.Phys.104(1986)605
\item[[35]]D.Kutasov, N.Seiberg, More Comments on String Theory on $AdS_3$,hep-th/9903219, JHEP 9904 (1999) 008
\item[[36]]Zamolodchikov AB and Fateev V A 1985 Sov.Phys. -JETP ${\bf62}$ 215\\D.Kastor, E.Martinec and Z.Qui, Phys.Lett.B200(1988)134\\D.Gepner, Nucl.Phys.B 290(1987),10\\ Belavin A A, Polyakov A M and Zamolodchikov A B 1984 Nucl. Phys.Lett.B ${\bf214}$ 333
\item[[37]]F Wilczek, Fractional statistics and anyon superconductivity, World scientific, singapore (1990)\\ S Forte, Mod.Phys Lett A6,(1991)3153, Mod Phys 64 (1992) 193.\\ J.Douari Anyons; statistiques fractionnaires et groupes quantiques, PhD Thesis, Faculty of Sciences, Rabat university (2000).
\item[[38]]B.Zweibach, Lecture Notes on string field theory delivered at the spring school on string theory (2000) ICTP Trieste Italy\\ A.Sen and  B.Zweibach hep-th/9912249 JHEP0003 (2000)002.
\item[[39]]A Giveon, M.Rocek hep-th/9904024 JHEP 9904(1999)019\\ R.Argurio, A.Giveon, Assaf  Shomer hep-th/0011046, The spectrum of N=3 String theory on AdS$_{3}\times G/H$
\end {enumerate}
\end{document}